\documentclass[dvipsnames]{aa}  
\usepackage[utf8]{inputenc}
\usepackage[T1]{fontenc}
\usepackage[dvipsnames]{xcolor}
\usepackage{txfonts}
\usepackage{graphicx}
\bibpunct{(}{)}{;}{a}{}{,}
\usepackage[colorlinks=true, linkcolor=blue, filecolor=blue, citecolor=blue, urlcolor=blue ]{hyperref}
\usepackage{xcolor}

\DeclareUnicodeCharacter{2212}{-}
\DeclareUnicodeCharacter{2032}{-}
\DeclareRobustCommand{\VAN}[3]{#2}
\let\VANthebibliography\thebibliography
\def\thebibliography{\DeclareRobustCommand{\VAN}[3]{##3}\VANthebibliography}
\begin{document} 
\titlerunning{ Lithium rich star in M67}
\title{Lithium rich stars in M67: Searching for hot companions through UV observations}
\author{S.~Rani\inst{\ref{oaa}}\thanks{E-mail: sharmila.rani@inaf.it},
        E.~Pancino\inst{\ref{oaa}},
        D.~Alvarez Garay\inst{\ref{oaa}},
        C.~Watts\inst{\ref{iia}},
        A.~Avdeeva\inst{\ref{oaa}},
        M.~Echeveste\inst{\ref{oaa}},
        E. I. Leitinger\inst{\ref{oaa}, \ref{unibo}},
        S.~Nedhath\inst{\ref{oaa}, \ref{unifi}},
        E. Reggiani\inst{\ref{oaa}, \ref{unifi}},
        N.~Sanna\inst{\ref{oaa}}, 
        S.~Saracino\inst{\ref{oaa}},
        L.~Steinbauer\inst{\ref{oaa}, \ref{unifi}},
        A.~Turchi\inst{\ref{oaa}}
        }

\institute{INAF - Osservatorio Astrofisico di Arcetri, Largo E. Fermi 5, 50125 Firenze, Italy\label{oaa}
\and Indian Institute of Astrophysics, Koramangala, Bangalore 560034, India\label{iia}
\and Dipartimento di Fisica e Astronomia, Università di Firenze,  Via G. Sansone 1, 50019 Sesto Fiorentino, FI, Italy\label{unifi}
\and Dipartimento di Fisica e Astronomia, Università degli Studi di Bologna, Via Gobetti 93/2, I-40129 Bologna, Italy\label{unibo}}
\authorrunning{S. Rani et al.}   
\date{Received: ---}

 
  \abstract
   {Lithium-rich (Li-rich) stars are a rare class of objects characterized by an abnormally high abundance of Li in their atmospheres. Although several explanations, including binary interactions, have been proposed, their origin remains uncertain.
   }
   {The main objective of this study is to search for hot companions to the known Li-rich stars in M67, characterize these stars, and evaluate the binary hypothesis as a possible explanation for their enhanced Li abundance.}
   {We used archival ultraviolet (UV) photometric data from several space missions, such as the UVIT/{\it AstroSat}, {\it GALEX}, and Swift/UVOT, in combination with optical and infrared photometry to create multiwavelength spectral energy distributions (SEDs), to search for possible hot companions to Li-rich stars. Further, we employed high-resolution UVES archival spectra to find any chemical signatures of past binary interactions.}
   {Our study reveals that two known Li-rich stars in M67, WOCS\,6008 and WOCS\,2003, exhibit a significant far-UV (FUV) excess relative to a single star SED fit, attributed to the presence of a companion in the former and to stellar activity in the latter. Thus, we report, for the first time, the detection of a hot companion to a Li-rich star in M67, offering direct evidence of the role of binary interactions in explaining its Li enrichment. The hot companion is found to be an extremely low mass (ELM) white dwarf (WD) of mass $\sim 0.16\pm0.01 M_{\odot}$, consistent with its kinematic mass and age of 2.5 Gyr. We did not find any enhancement of Na or s-process elements, nor any depletion of C or Mg in WOCS\,6008, suggesting that mass transfer during the asymptotic giant branch or red giant branch stage is unlikely. We suggest this star is the accretor component of a binary system that underwent early Case B mass transfer, as supported by its derived elemental abundances.}
   {We report the first direct detection of the UV flux from a compact companion to a Li-rich star, WOCS\,6008. Our results support previous suggestions that stellar rotation is the most favored mechanism accounting for the Li enrichment in this post interaction system. Overall, the available evidence supports the case for binary interactions as one of the channels to form Li-rich stars.
   }

   \keywords{open clusters and associations: individual: M67 – Stars: chemically peculiar – binaries: spectroscopic - Stars: formation}

   \maketitle
%

\section{Introduction}
\label{sec:intro}
Lithium-rich (Li-rich) stars represent a class of chemically peculiar objects whose atmospheres exhibit anomalously high Li abundances $A(\mathrm{Li})$ compared to stars at similar evolutionary stages. These rare objects \citep[$\sim1~\%$ of all stars,][]{2016ApJ...819..135K, 2016MNRAS.461.3336C, 2020MNRAS.494.1348D} were first discovered in the giant phase by \cite{1982ApJ...255..577W} with $A(\mathrm{Li}) = 3.2~\mathrm{dex}$. Since then, they have been detected across all evolutionary phases and in diverse environments, including the field \citep{2016MNRAS.461.3336C, 2018ApJ...852L..31L, 2020MNRAS.494.1348D, 2024A&A...689A..89M}, globular clusters \citep[GCs;][]{ 2016ApJ...819..135K, 2019A&A...623A..55M, 2020A&A...639L...2S}, open clusters \citep[OCs;][]{2014A&A...564L...6M, 2023A&A...674A.157T}, and dwarf spheroidal galaxies \citep{2012ApJ...752L..16K}. The existence of these Li-rich stars is particularly puzzling, because 
canonical stellar evolution theory predicts Li depletion, first as the star evolves off the main sequence (MS) to the red giant branch (RGB) due to mixing caused by the first dredge-up and later during subsequent evolutionary mixing episodes throughout the RGB evolution \citep{1967ApJ...147..624I, 2026ARA&A..64..587C}. Their origin still lacks a proper explanation and requires further exploration.

Several theories have been proposed in the past to explain the high Li content in these exotic stars located at different evolutionary phases, from the MS to the post-asymptotic giant branch (post-AGB). To date, the formation routes of such stars include internal production of Li through the Cameron-Fowler mechanism \citep[CF;][]{1971ApJ...164..111C} or contamination of Li-rich material from external sources such as AGB or RGB stars that have already passed through the Li production process \citep{1999ApJ...510..217S}, accretion of ejecta from novae or supernovae enriched in Li \citep{1978ApJ...222..600S, 1995ApJS..101..181W, 2025A&A...699A.171M}, engulfment of planets or substellar systems \citep{1999MNRAS.308.1133S, 2016A&A...587A..66D, 2026arXiv260521407K}, and tidal spin-up from a binary companion \citep{2019ApJ...880..125C}. Additionally, recent studies have suggested that the mechanism responsible for the Li-enrichment in stars depends on its specific evolutionary stage \citep{2019ApJ...880..125C, 2020NatAs...4.1059K, 2023ApJ...944L...5M}.

As mentioned above, binarity is one of the scenarios invoked to explain Li enhancement in these stars. This scenario can explain high Li content in three ways: the first is accretion of Li enhanced matter from a companion such as an intermediate-mass AGB or RGB star. In the case of an AGB star, the CF mechanism can be triggered to produce Li, whereas an RGB star produces fresh Li through extra mixing processes such as thermohaline mixing \citep{2007A&A...467L..15C}, magneto-thermohaline mixing \citep{2009ApJ...696.1823D}, and rotation-induced mixing \citep{1992A&A...255..191C, 1995ApJ...453L..41C}. 
The second mechanism is mass transfer from a high-mass companion to a low-mass companion that can increase the latter's angular momentum, which in turn leads to high rotation that induces mixing in the star, bringing Li up to the stellar surface \citep{Sayeed2023}. The third mechanism was proposed by \cite{1994A&A...288..829Z} to account for the enhancement of Li in close binary systems. They suggested that tidal interactions may contribute to the preservation of the Li abundance from the MS turn-off stage. Tidally locked binary systems on the ZAMS also experience tidal torques, which tend to synchronize their rotational and orbital motions. As a result, these binary stars exchange less angular momentum and thus retain more of their original Li.

In this work, we present multiwavelength photometric analysis of two previously identified Li-rich stars (WOCS\footnote{WOCS (WIYN open cluster study) ID is taken from \cite{2015AJ....150...97G} and we refer to this ID in the entire text.}\,2003 and WOCS\,6008), members of the extensively studied OC M67. The primary aim is to search for unseen hot companions to the Li-rich stars using ultraviolet (UV) photometry and to constrain their possible formation pathways based on both photometric and spectroscopic data. Additionally, we provide high-resolution spectroscopic abundance analysis for one of the two Li-rich stars (WOCS\,6008), which is a subgiant and classified as a spectroscopic binary \citep[SB1;][]{1990AJ....100.1859M, 2015AJ....150...97G, 2021AJ....161..190G}, to look for possible signatures of AGB or RGB pollution.

The organization of the paper is as follows: In Sect.~\ref{sec:data}, we present the targets, the data sources, and the data reduction procedure. In Sect.~\ref{sec:sed_ana}, we discuss the SED analysis to obtain the stellar parameters, followed by the characterization of hot companions in Sect.~\ref{sec:hot_comp}. Sect.~\ref{sec:variable} provides a detailed description of photometric variability. In Sect.~\ref{sec:chem}, we present the spectroscopic abundance analysis results for the binary system WOCS\,6008. The discussion of the results is presented in Sect.~\ref{sec:dis}. Finally, we summarize and conclude our findings in Sect.~\ref{sec:summary}.


\begin{figure}
    \centering
    \includegraphics[width=1.1\columnwidth]{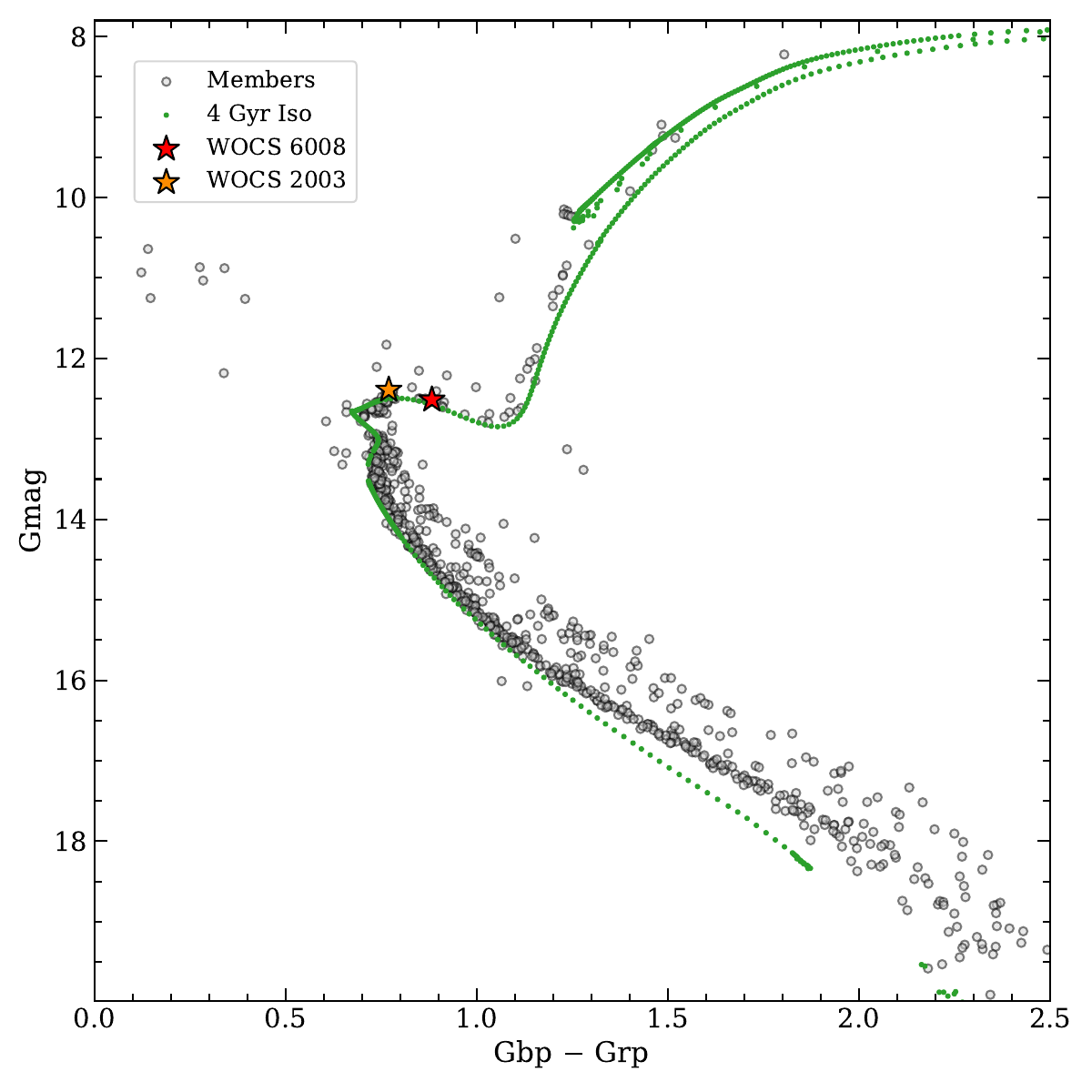}
    \caption{\textit{Gaia} CMD of M67 showing the location of the two known Li-rich stars marked with red and orange star symbols. The gray points represent the stars with proper motion membership probability more than 50$\%$. The green dotted line represents the MIST isochrone with solar metallicity and cluster age \citep{2016ApJ...823..102C, 2018ApJS..234...34P}.}
    \label{fig:cmd}
\end{figure}

\begin{table*}
  	\begin{center}
	\caption{Basic properties of the targets analyzed in this study. The first three columns indicate the WOCS ID, RA, and DEC information of the sources, followed by membership probabilities in terms of PM and RV, orbital period, eccentricity, projected rotational velocity, and rotational period.}
	\label{tab1}
    \large
   \resizebox{2.0\columnwidth}{!}{%
\begin{tabular}{cccccccccc} 
		\hline
		\hline
		 WOCS\,ID & \textit{Gaia} DR3 ID &RA  & Dec & P$_{PM}$ & P$_{RV}$ & Orbital Period & eccentricity  & vsini &  Rotational Period \\
         & & (deg) & (deg) & & & (days) &  & (km s$^{-1}$) & (days)\\
		\hline
		6008 & 604916559907415040 & 132.899993 & 11.775989 & $89^{a}$ & $98^{b}$ & $31.78 \pm 0.0006^{c}$ & $0.66 \pm 0.01^{c}$ & $6.0\pm1.0^{d}$ & $6.54\pm0.1^{f}$\\ 
        2003 & 604920683076006272 & 132.829186 & 11.834918 & $100^{a}$ & $98^{b}$ & $7.65 \pm 0.0001^{c}$ & $0.008 \pm 0.004^{c}$ & $9.48\pm0.79^{e}$ & $7.88\pm0.2^{f}$ \\ 
		\hline 
	\end{tabular}
  }
  
  {\footnotesize \textbf{References:}   a)~\cite{2023A&A...673A.114H},
  b)~\cite{2015AJ....150...97G},
  c)~\cite{2021AJ....161..190G},
  d)~\cite{2006A&A...451..993C},
  e)~\cite{2023A&A...676A.129H} 
  f)~\cite{2023A&A...672A.159G}.}
	\end{center}
\end{table*} 

\section{Target Selection and Data Reduction}
\label{sec:data}
\subsection{Li-rich Stars in M67}
To date, two Li-rich stars have been identified in the OC M67. The first, WOCS\,2003 \citep[S1045,][]{1977A&AS...27...89S}, was discovered by \cite{1994ApJ...434L..71D} and \cite{1997A&A...325..535P} with an $A(\mathrm{Li})$ value of $3.0~\mathrm{dex}$. WOCS\,2003 is a short-period tidally locked binary (SPTLB) situated in the Li peak region of M67, and its Li abundance is much higher than the one expected in the Li-peak ($A(\mathrm{Li}) = \sim 2.5~\mathrm{dex}$). It is classified as a double-lined spectroscopic binary (SB2) with an orbital period of 7.65 days, and an eccentricity of 0.014$\pm$0.006 \citep[S1045,][]{1990AJ....100.1859M}. \cite{1994ApJ...434L..71D} suggested that this system is composed of two components with similar temperatures located near the MS turn-off. Later, \cite{2006A&A...451..993C} reported the detection of a second Li-rich star, WOCS\,6008 (S1242), located on the subgiant branch (SGB) of M67. WOCS\,6008 is known to exhibit the highest Li abundance ($A(\mathrm{Li}) = 2.7~\mathrm{dex}$) ever derived for M67 stars evolving off the MS. \cite{2012A&A...541A.150P} compiled literature Li abundances of M67 stars from the MS to the RGB. Their compilation shows that most stars of similar mass (1.29--$1.35\,\mathrm{M_{\odot}}$) that have already left the MS have much lower Li abundances, with a median of $A(\mathrm{Li}) = 1.10$\,dex (see their Fig.~2). WOCS\,6008 therefore stands out with a Li overabundance of $1.60 \pm 0.61$\,dex relative to stars in the same evolutionary phase.


\begin{figure}
    \centering
    \includegraphics[width=0.9\columnwidth]{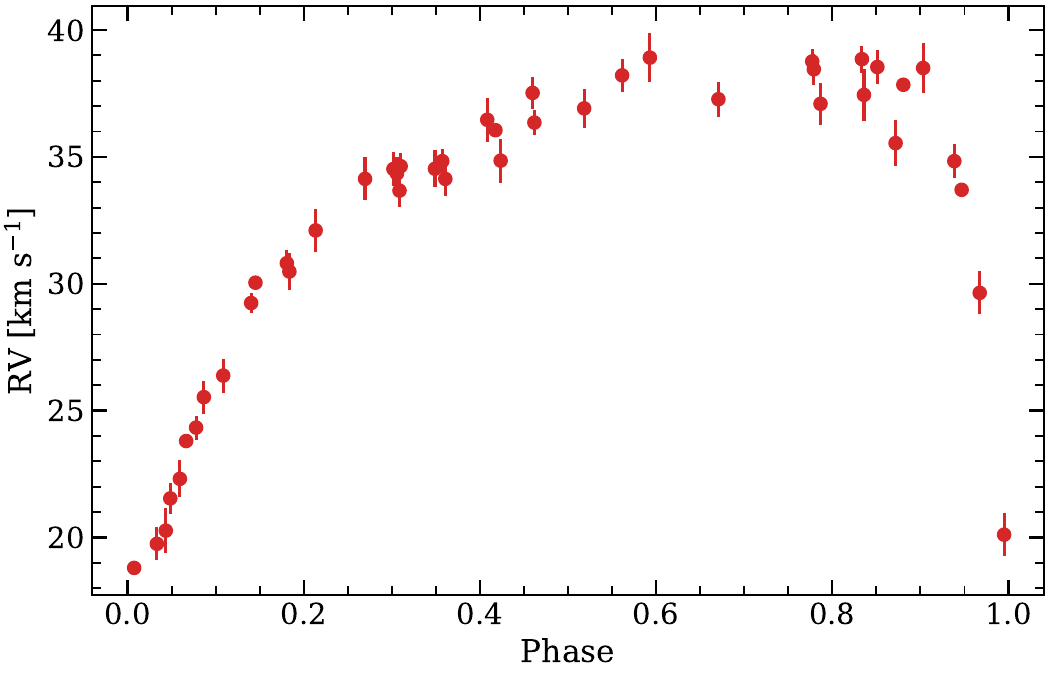}
    \caption{Heliocentric RVs as a function of orbital phase for WOCS\,6008, taken from \cite{2015AJ....150...97G}.}
    \label{fig:rv_curve}
\end{figure}

Figure~\ref{fig:cmd} shows the location of both stars in the {\it Gaia} optical color magnitude diagram (CMD) where gray points correspond to the stars with proper motion membership probability more than 50$\%$ \citep{2016A&A...595A...1G, 2023A&A...674A...1G}. Furthermore, \cite{2011A&A...527A..94C} measured Li abundances of 27 evolved stars (turn-off, subgiant, and giant) in this cluster and demonstrated that the Li abundances follow the now well-known evolutionary pattern from the turn-off to the RGB. None of the stars in their studied sample exhibited enhanced Li. 

\par
In this work, we mainly focus on WOCS\,6008, while WOCS\,2003 has been used for comparison. The basic information and kinematic properties of these systems from the literature, including orbital period, eccentricity, and projected rotational velocity, are presented in Table~\ref{tab1}.

\begin{figure*}
    \centering
    \includegraphics[width=0.5\columnwidth]{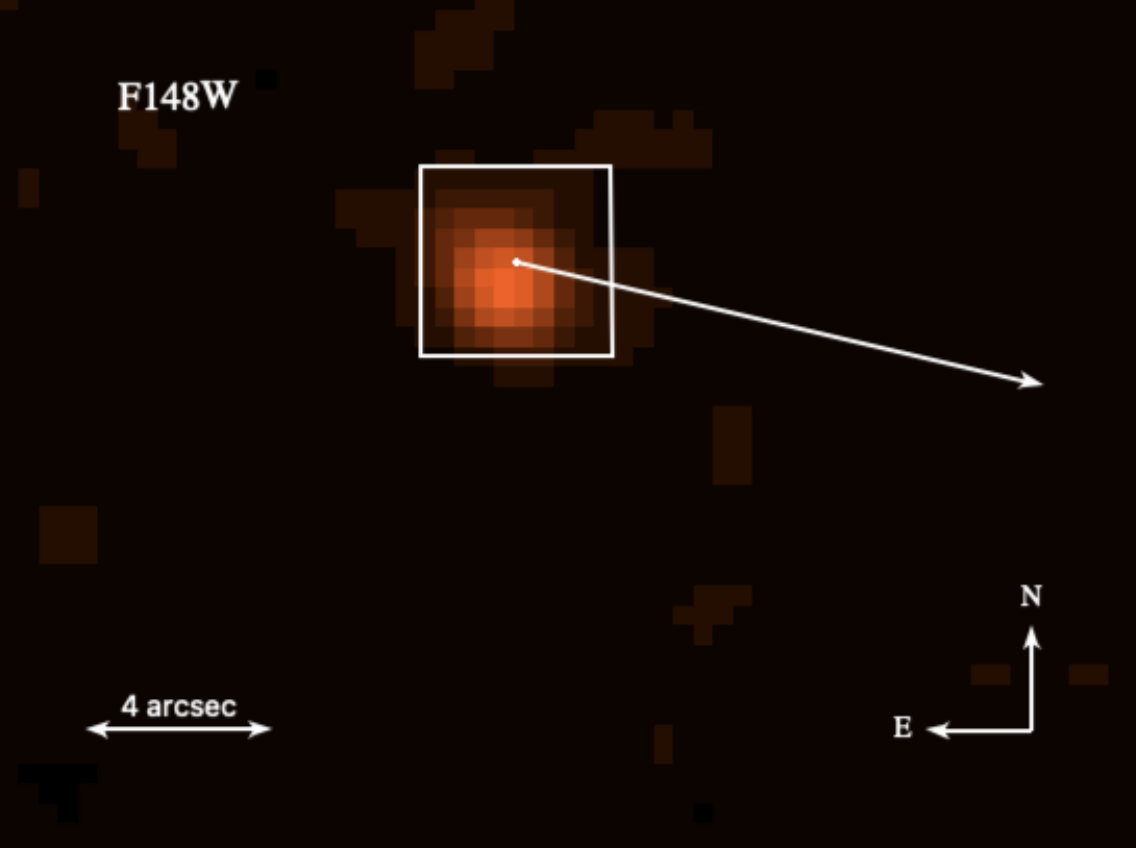}
\includegraphics[width=0.5\columnwidth]{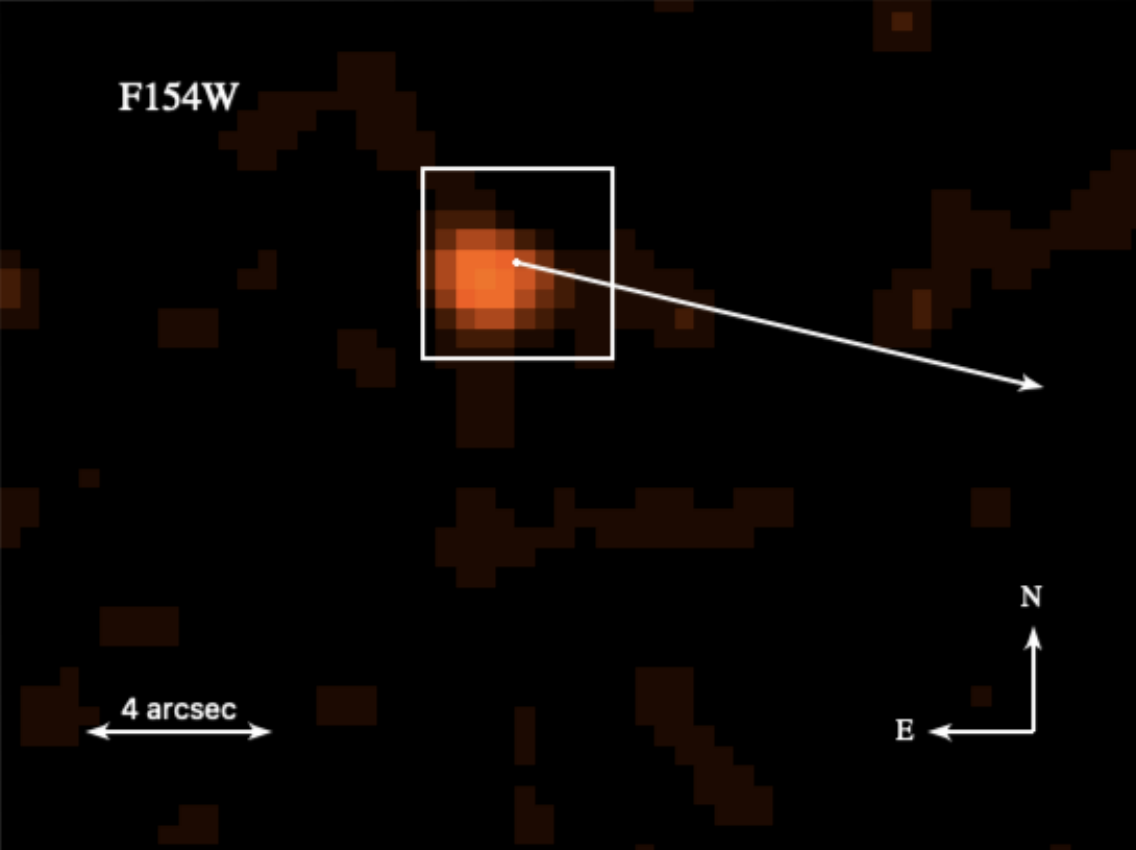}
\includegraphics[width=0.5\columnwidth]{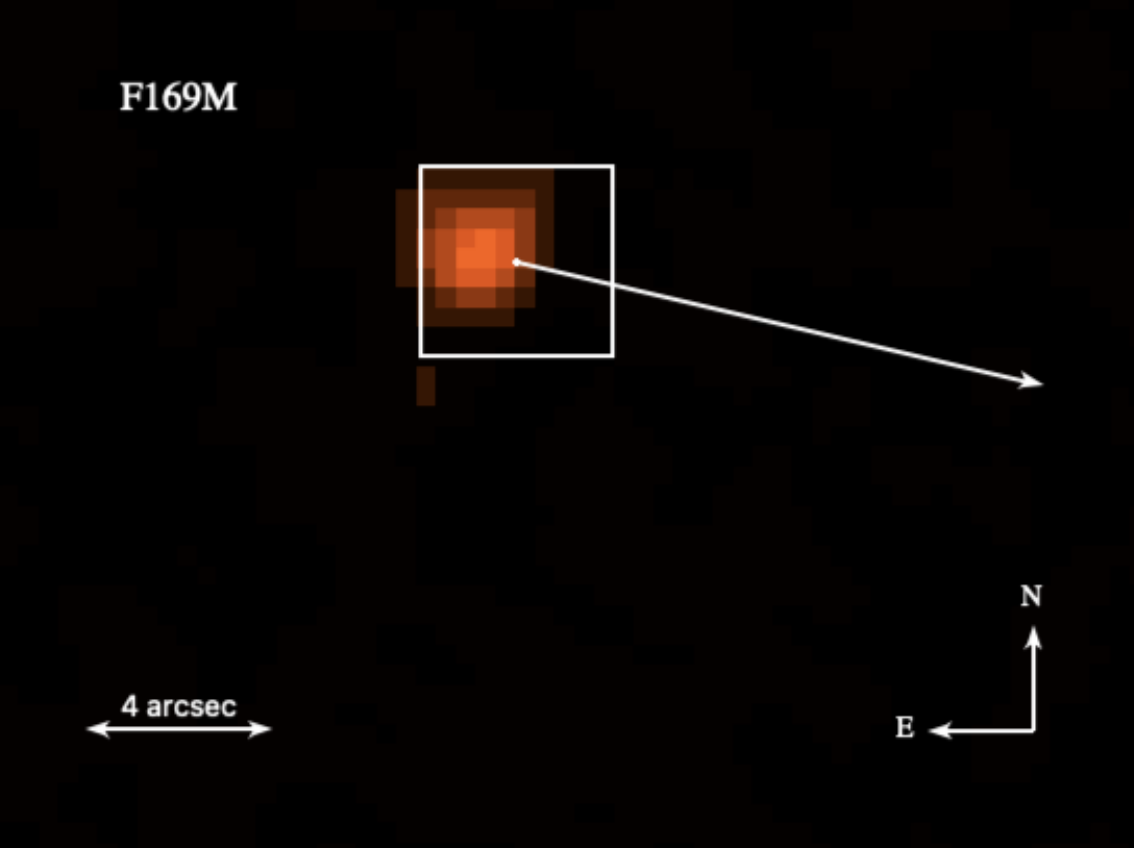}
\includegraphics[width=0.5\columnwidth]{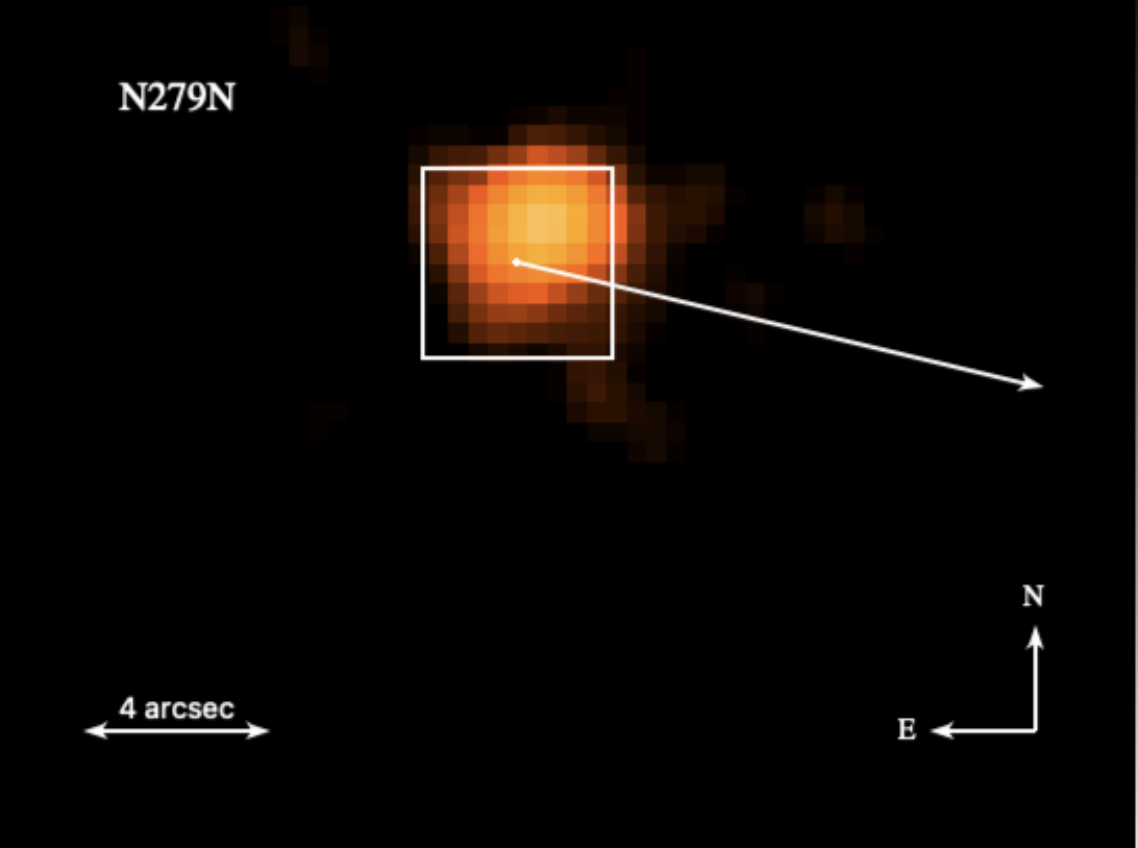}
\includegraphics[width=0.5\columnwidth]{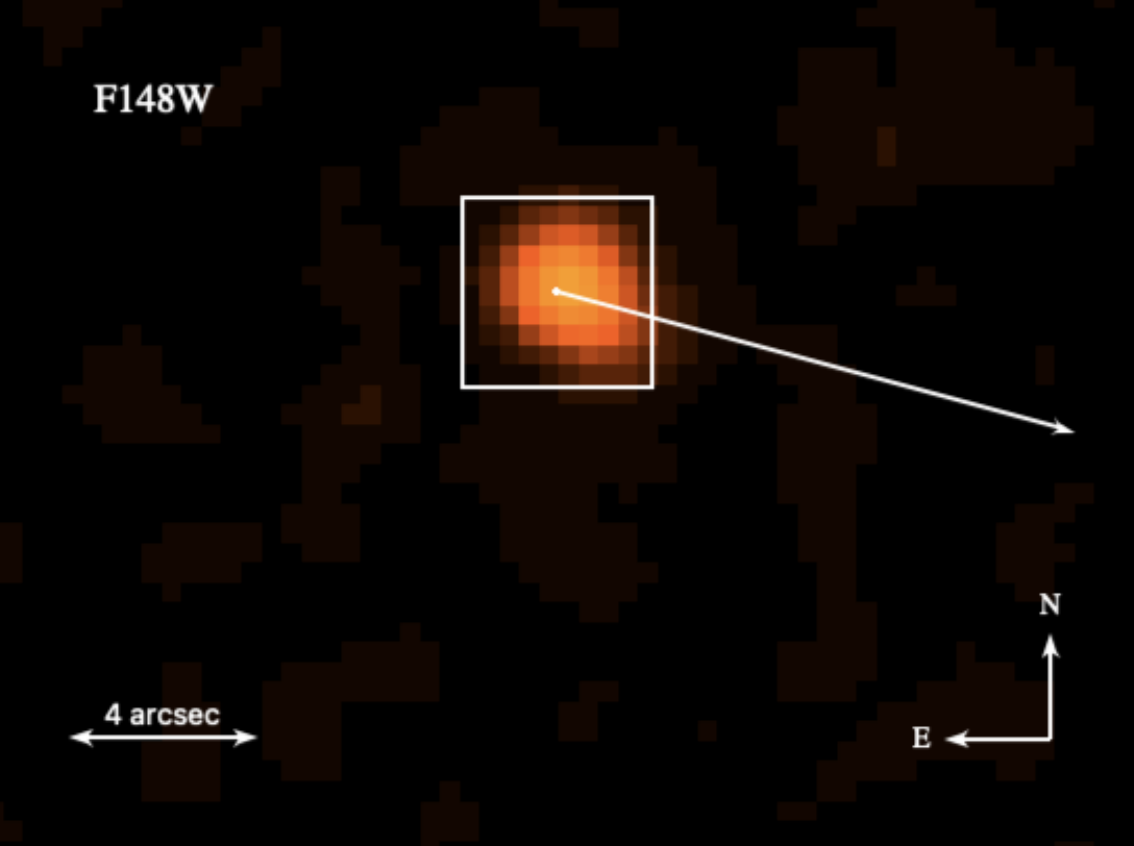}
\includegraphics[width=0.5\columnwidth]{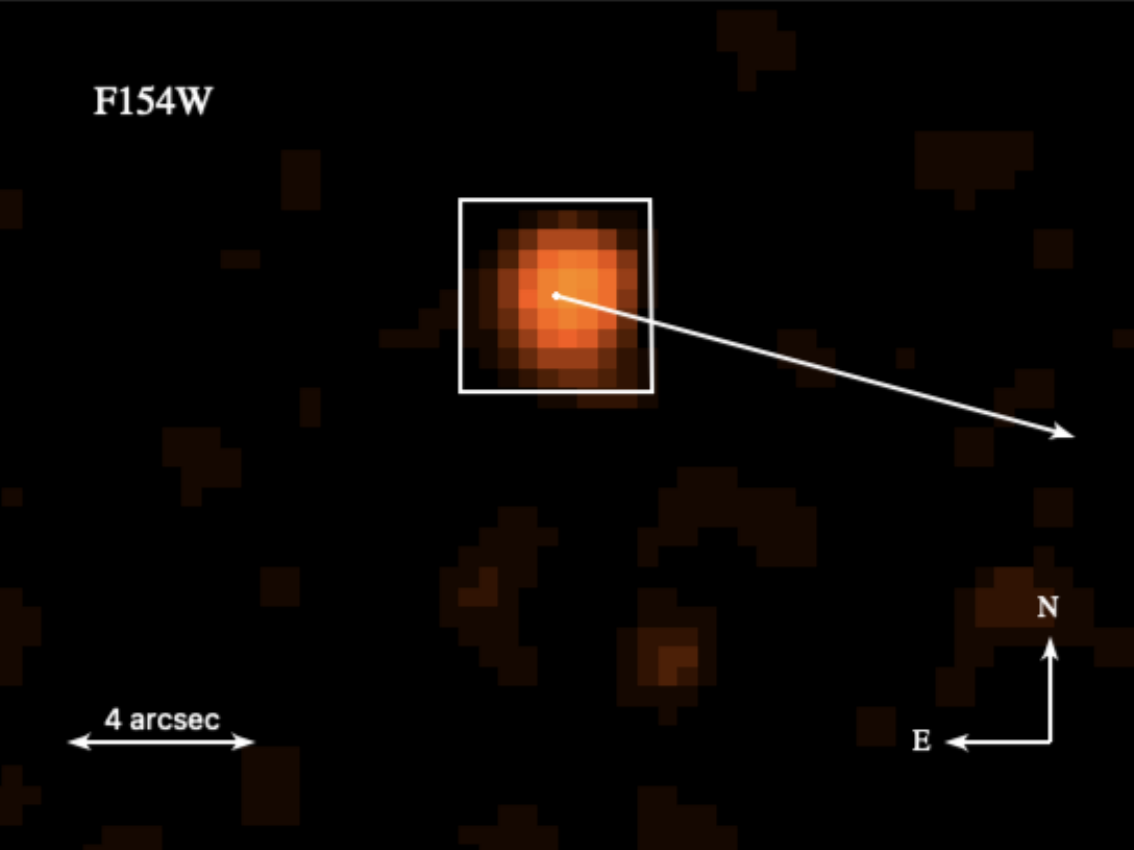}
\includegraphics[width=0.5\columnwidth]{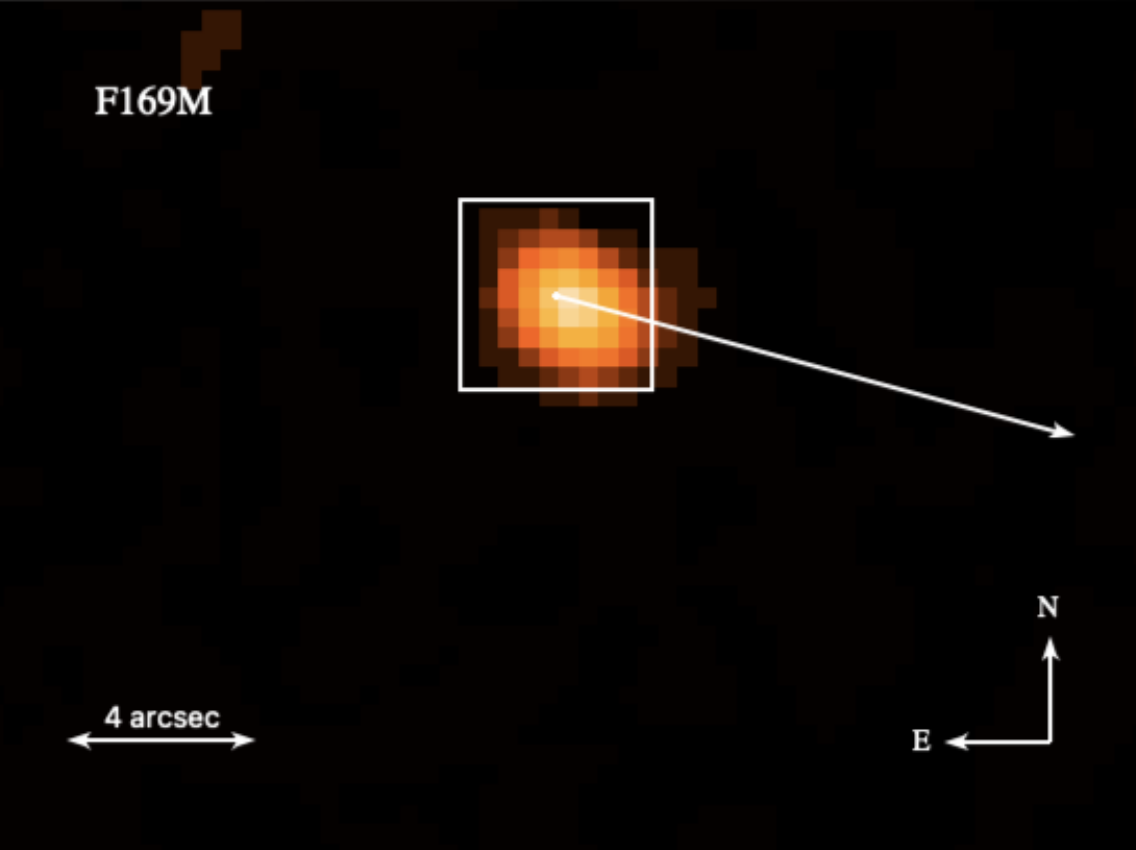}
\includegraphics[width=0.5\columnwidth]{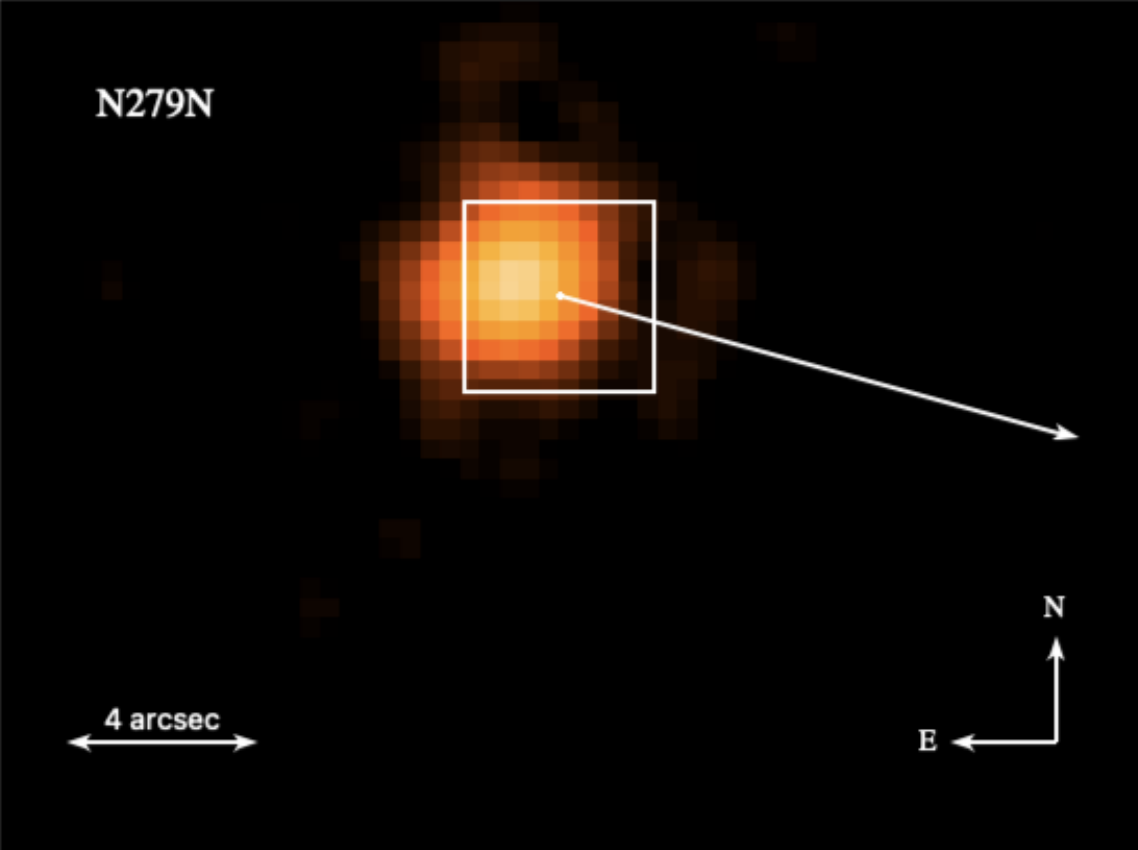}
    \caption{Detection of Li-rich stars, WOCS\,6008 (upper panel) and WOCS\,2003 (lower panel), illustrated in the different UVIT filters along with proper motion direction shown with the white downward arrow.}
    \label{fig:location}
\end{figure*}

\subsection{UV Photometric Data}
\label{sec:uvdata}
Since WOCS\,6008 was previously identified as a single-lined spectroscopic binary \citep[SB1, S1242,][]{1990AJ....100.1859M}, we anticipate that it may have a UV-bright companion that is faint in the optical regime and, hence, difficult to identify in optical photometric and spectroscopic observations. The heliocentric RVs of WOCS\,6008 as a function of orbital phase are shown in Fig.~\ref{fig:rv_curve}. Here, we leverage UV photometric data to search for its potential hot companion to confirm its binary nature and shed light on its Li enrichment history and formation in this cluster. We used the spectral energy distribution (SED) fitting technique to model a binary system consisting of hot and cool components, thereby revealing the nature of the hot companion, as detailed in Sect.~\ref{sec:sed_ana}.

In order to construct the multiwavelength SEDs of stars under analysis, we primarily employed archival UV photometric data from the Ultra Violet Imaging Telescope (UVIT) aboard \textit{AstroSat}, India's first multiwavelength space observatory \citep{2017AJ....154..128T}. M67 observations were conducted with UVIT in April 2017 and December 2018 using NUV N279N and FUV F148W, F154W, and F169M filters. We used CCDLAB software \citep{2017PASP..129k5002P} to process L1 data from the \textit{AstroSat} data archive\footnote{\url{https://astrobrowse.issdc.gov.in/astro_archive/archive/Home.jsp}} to obtain science-ready images. CCDLAB corrects for satellite drift, flat-field, distortion, fixed-pattern noise, and cosmic rays. The orbitwise raw images are then aligned and combined to obtain a composite image with an increased signal-to-noise ratio (SNR). The log of observations of UVIT data is listed in Table 1 of \cite{2021MNRAS.507.2373P}. The full details of the UVIT instrument, its in-orbit performance, and calibration are described in \cite{2017AJ....154..128T, 2020AJ....159..158T}. We further performed PSF photometry on UV images using the DAOPHOT package in IRAF/NOAO \citep{1987PASP...99..191S}, followed by PSF, aperture, and saturation corrections to the PSF magnitudes to obtain the instrumental magnitudes. As UVIT uses the standard AB magnitude system similar to {\it GALEX}, we calibrated instrumental magnitudes into the AB magnitude system using zero-point (ZP) magnitudes reported in \cite{2017AJ....154..128T}. Thanks to the deep FUV observations taken with UVIT, WOCS\,6008 and WOCS\,2003 are detected in all UVIT FUV and NUV images, as shown in Figure~\ref{fig:location}. The PSF photometric errors on UVIT magnitudes of both stars are $\sim$0.2\,mag.

\begin{figure}
    \centering
    \includegraphics[width=\columnwidth]{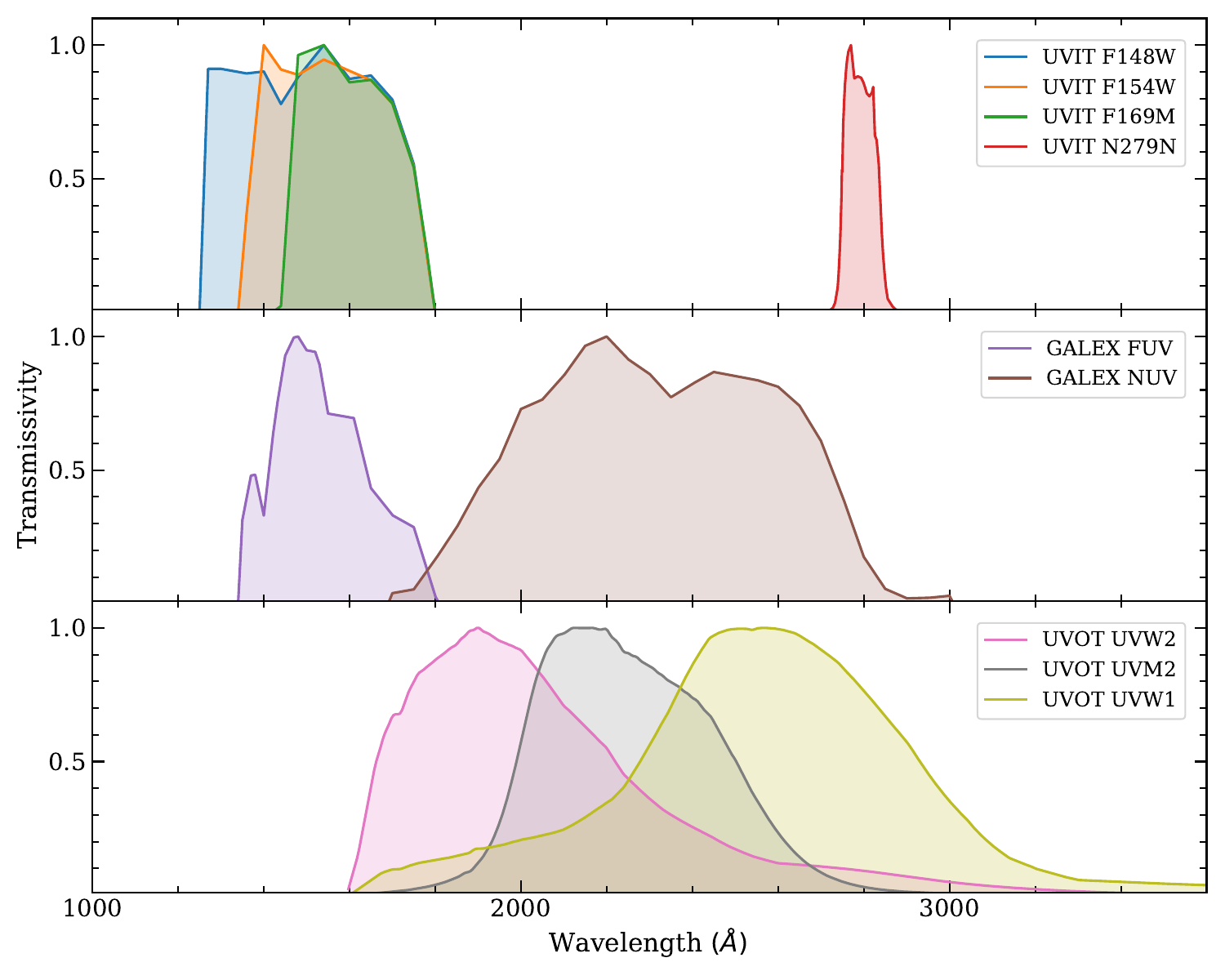}
    \caption{Transmission curves of UVIT (upper), {\it GALEX} (middle), and Swift/UVOT (lower) filters, retrieved from the SVO Filter Profile Service (\url{https://svo2.cab.inta-csic.es/theory/fps/}).}
    \label{fig:trans}
\end{figure}

In addition to the UVIT data, we also utilized archival UV photometric catalogs from other space missions such as {\it GALEX} in FUV and NUV passbands \citep{galexcatalog}, and Swift/UVOT in three NUV bands (uvw2, uvm2, uvw1) and one UV u band \citep{2019AJ....158...35S}. We show the transmission curves of all UV filters used in this work in Figure~\ref{fig:trans}. We further cross-matched UVIT-detected sources in various FUV and NUV filters with {\it GALEX} and Swift/UVOT catalogs using a cross-match radius of 1.1$\arcsec$, corresponding to UVIT spatial resolution. As our study focuses on the two Li-rich stars previously reported in the literature, we searched for their UV counterparts in the aforementioned catalogs.  Additionally, both are detected in {\it GALEX} FUV and NUV images and in all NUV filters of Swift/UVOT.

For the optical and infrared (IR) regions, we compiled archival data from several catalogs, including \textit{Gaia} Data Release 3 (DR3) \citep{2023A&A...674A...1G}, SDSS \citep{sdsscatalog}, Pan-STARRS1 \citep{2016arXiv161205560C}, APASS \citep{apasscatalog}, 2MASS \citep{2masscatalog}, and WISE \citep{allwisecatalog}. These archival data were combined with the UV data to construct a multiwavelength SED covering the full range from UV to IR (0.13–11.5 $\mu$m). The virtual observatory tool in the Virtual Observatory SED Analyzer (VOSA; \citealp{2008A&A...492..277B}) was employed to cross-identify the optical and IR counterparts of both targets using a matching radius of 2$\arcsec$. We assumed a distance of 865$\pm$18 pc \citep{2021MNRAS.504..356D}, solar metallicity, and reddening value of 
E(B$-$V) = $0.04 \pm 0.01\rm mag$ \citep{2021MNRAS.504..356D}.

\subsection{Optical Spectroscopic Data}
\label{sec:specdata}
Chemical peculiarities can be used to identify the evolutionary phase in which the mass transfer, if any, happened in a binary system. A possible signature to identify an AGB companion to a Li-rich star could be the detection of enhanced s-process elements on the surface of the Li-rich star if mass transfer happened after the thermally pulsing phase \citep{1999ARA&A..37..239B}. AGB stars can also be responsible for the variation in the light element abundances \citep{2013MNRAS.431.3642V}. In contrast, no such signature is expected from RGB or MS stars as companions. For this purpose, we employed archival high-resolution spectroscopic data of WOCS\,6008 obtained with the UVES spectrograph \citep{2000SPIE.4008..534D} mounted on the VLT (UT2). Observations were carried out on 5 January 2004 with the blue and red arms, covering wavelength ranges of 476-580 nm and 582-683 nm, respectively, at a spectral resolution of $\sim$ 50000. The signal-to-noise ratio (SNR) per pixel in the blue and red regions of the spectrum is $\sim$130 at 6140 $\AA$ and $\sim$150 at 5200 $\AA$, respectively. 

We retrieved spectra from the ESO archive, which are reduced using the dedicated UVES ESO pipeline\footnote{\url{https://archive.eso.org/scienceportal/home}} that includes bias subtraction, flat-field correction, spectral extraction,
wavelength calibration, sky subtraction, and order merging. We used the iSpec tool \citep{2019MNRAS.486.2075B} for continuum normalization of the wavelength-calibrated spectra. The radial velocity (RV) of WOCS\,6008 was determined using the cross-correlation method implemented in iSpec and was estimated to be 36.42 $\pm$ 0.14 km/s, in excellent agreement with previous estimates \citep{ 2015AJ....150...97G, 2021AJ....161..190G, 2023A&A...676A.129H}. We used the Sun spectrum as a reference template \citep{2011ApJS..195....6W}. Further, we used these spectra to measure the abundance ratios of interesting elements 
using the spectral synthesis technique described in detail in Sect.~\ref{sec:chem}.

\section{SED Analysis}
\label{sec:sed_ana}
 We constructed the SEDs of both Li-rich stars using the broadband photometry described in Sect.~\ref{sec:uvdata} spanning wavelengths from the FUV to IR. Using VOSA, we constructed and fitted the observed SEDs with theoretical Kurucz models \citep{1997A&A...318..841C, 2003IAUS..210P.A20C}. VOSA performs a reduced chi-square ($\chi^2_\mathrm{red}$) minimization to determine the best-fit stellar parameters, including effective temperature (T$_{\mathrm{eff}}$), luminosity (L), and radius (R). To compute the predicted synthetic flux corresponding to each photometric data point used in the observed SED, VOSA convolves the model spectra with filter transmission curves. All observed fluxes are corrected for interstellar reddening, using the Fitzpatrick reddening law with E(B$−$V) = $0.04 \pm 0.01 \rm mag$ \citep{1999PASP..111...63F, 2005ApJ...619..931I}. To indicate a good fit, fractional residuals between the observed and synthetic fluxes should be minimized. Additionally, VOSA calculates two modified $\chi^2_\mathrm{red}$ parameters, called the visual goodness of fit (Vgf) and Vgf$_b$, to assess the overall fit when observational flux errors are very small. To ensure a reliable SED fit, we use Vgf$_b$ < 15 following the criterion adopted by \cite{2021MNRAS.506.5201R}. The scaling factor $\rm M_d = (R/D)^2$ is used to calculate the radius of a star for a given distance D (pc). To estimate uncertainties in the derived parameters, VOSA utilizes the Markov Chain Monte Carlo (MCMC) approach. In case estimated errors using the MCMC approach are less than half of the model grid size, it will report half of the grid size as a final error in the stellar parameters. More details on the SED fitting method adopted here can be found in \citet{2023ApJ...945...11R}.

\subsection{Single Model Fit}
\label{sec:singmodfit}
First, we fitted the observed SEDs, incorporating all data points, with a single model SED in order to assess the overall fit. For this purpose, we used the Kurucz atmospheric models ($\rm [\alpha/Fe]$ = 0) that offer a grid of stellar parameters such as T$_{\mathrm{eff}}$, metallicity [Fe/H], and surface gravity ($\log g$). To fit the observed SEDs of the stars, we treated T$_{\mathrm{eff}}$ and $\log g$ as free parameters, while fixing the metallicity to [Fe/H] = 0.0, consistent with the cluster's metallicity. We adopted a T$_{\mathrm{eff}}$ range of 3,500--10,000~K and a $\log g$ range of 3.5--5~dex in the Kurucz models, since the two stars considered for analysis are found along the SGB and MS turn-off in M67. Both stars show a significant deviation in flux ($>$50\%) from the model at shorter wavelengths (FUV region), and consequently, a single model fit does not reproduce the observed SEDs. We further repeated the single model SED fit by excluding data points with wavelengths shorter than 3000 \AA, and including only optical and IR data points to obtain an estimate of best fit stellar parameters for the cooler component. Using {\it GALEX} photometric data in the FUV and NUV bands, \cite{2018MNRAS.481..226S} studied UV stellar populations of M67 to identify stars showing UV excess, including WOCS\,6008. However, they have not further characterized the targets with UV excess due to only one data point in the FUV regime. Our derived stellar parameters for the cooler component of WOCS\,6008 are in excellent agreement with their estimate. 

\subsection{Other Potential Sources of UV excess}
\label{sec:activity}
There are several factors that contribute to the excess flux in the UV region in stars. Main sources include the presence of a potential hot companion \citep{2008ApJ...689.1274S, 2016MNRAS.461.3036O}, chromospheric activity \citep{2013MNRAS.431.2063S}, coronal activity \citep{2015ApJ...810...77S}, and accretion onto a hot compact object \citep{2011ApJ...740L..39S}. Both Li-rich stars display significant UV excess, as illustrated in Figure~\ref{sed_fit}, and have been reported to be active in the literature. From the detection of emission features in the Ca II H \& K lines, both are identified as chromospherically active stars \citep{1998A&A...336..902P, 1999A&A...347..866V}. However, chromospheric activity cannot account for the FUV excess observed in these targets, as they exhibit an excess of more than 50$\%$, and chromospheric activity is known to cause excess flux primarily at NUV wavelengths \citep{2016MNRAS.463.2125P, 2020A&A...642A.228W, 2020AJ....160...12D, 2023MNRAS.518.4579P}. In the FUV regime, UV excess detected in FGK-type stars is generally indicative of a hot compact companion \citep{2016MNRAS.463.2125P,  2018PASA...35...37S, 2020A&A...642A.228W, 2023AJ....165...70C, 2023MNRAS.518.4579P}. Furthermore, some studies suggest that chromospheric activity among red giants may also result from the presence of a companion in a binary system \citep{2018PASA...35...37S, 2025AJ....169..309D}. 

Based on X-ray studies conducted using ROSAT, Chandra, and XMM-Newton observations of M67, both targets are reported as X-ray sources \citep{1998A&A...339..431B, 2004A&A...418..509V, 2015MNRAS.452.3394M}. In fact, WOCS\,2003 is found to be an RS Canum Venaticorum (RS CVn) variable based on both chromospheric and X-ray activity. It has been demonstrated in previous studies that X-ray activity results in emission lines within the transition region (TR) in the FUV regime, producing FUV excess that can mimic the presence of a hot companion \citep{1981ApJ...247..545A}. Since WOCS\,2003 is an SB2 system with both components on the MS and having similar temperatures and masses \citep{1994ApJ...434L..71D, 2021AJ....161..190G}, the detected UV excess is attributed to the X-ray activity of the system. Hence, the possibility of the presence of a potential hot companion in WOCS\,2003 is ruled out. Moreover, the caveat of SED fitting is that it cannot be used for the disentanglement of components with similar temperatures. Therefore, characterization of both components in WOCS\,2003 using this technique is not possible. 

In contrast, WOCS\,6008 is an SB1 system with large eccentricity; the coronal and chromospheric activity observed in this system is likely driven by rapid rotation resulting from tidal interactions at periastron or by a low-mass active M dwarf. As M dwarfs span a mass range of 0.1-0.6 $M_{\odot}$, it is quite possible that an active M dwarf (i.e., M5V) with a mass of about 0.15 $M_{\odot}$, similar to the kinematic mass of the companion, could be a companion that mainly contributes to the excess FUV flux seen in this system. To test this possibility, during our SED fitting, we fit several spectra of active early M dwarf stars from the MUSCLES survey \citep{2025ApJ...978...85W}\footnote{\url{https://archive.stsci.edu/hlsp/muscles}}. As an example, Fig.~\ref{fig:sed_mdwarf} (Appendix~\ref{app:appB}) shows a composite SED fit combining a Kurucz model spectrum of the SGB with the observed spectrum of an active M5 dwarf. However, the FUV flux of these stars was always several orders of magnitude too low to explain the observed UV flux, indicating that the activity alone is not sufficient to contribute enough UV excess to account for our observations of the low mass WD. Thus, we conclude that the possibility of an M dwarf as a companion is unlikely.


As the detected FUV excess in WOCS\,6008 could be contaminated by the one expected from X-ray activity, we further estimated and subtracted the FUV contribution attributable to such activity. For this purpose, we selected the observed International Ultraviolet Explorer (IUE) FUV spectrum of an active single star with similar stellar parameters and activity levels, and then convolved it with the FUV filters to estimate the corresponding fluxes. Furthermore, to isolate the photospheric contribution to the FUV emission of G stars, we followed the procedure described by \cite{2013MNRAS.431.2063S} and found that it is negligible below 1800 $\AA$. The estimated FUV flux solely due to activity ($\sim 10^{-18}\,\textrm{erg}\,\textrm{cm}^{-2}\,\textrm{s}^{-1}\,\textrm{\AA}^{-1}$) is two orders of magnitude too low to explain the observed FUV flux. Nevertheless, to stay on the safe side, we subtracted the activity contribution from the observed FUV fluxes in different passbands in WOCS\,6008. Our conclusion is that most of the observed FUV flux is likely caused by the presence of a hot companion. 

 \subsection{Composite Model Fit}
\begin{figure*}
    \centering \includegraphics[width=\columnwidth]{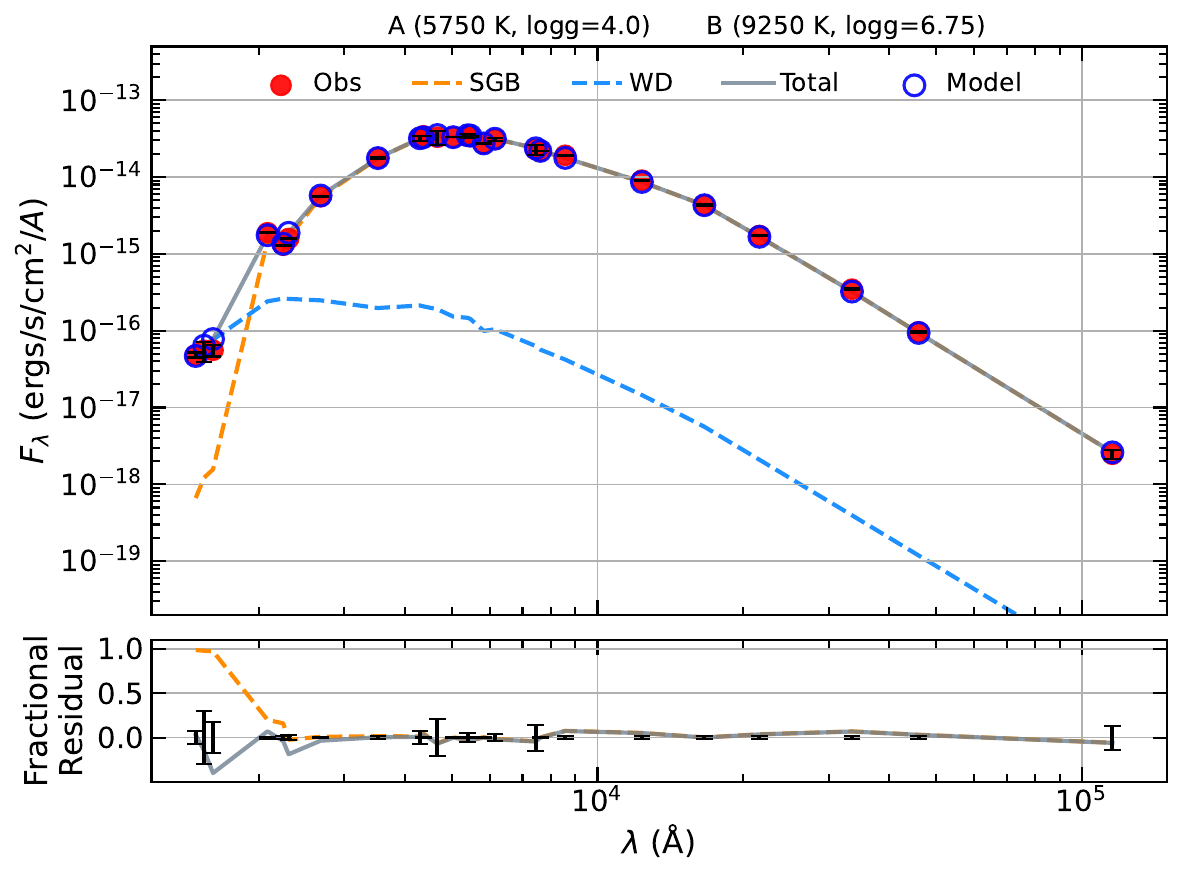}
\includegraphics[width=\columnwidth]{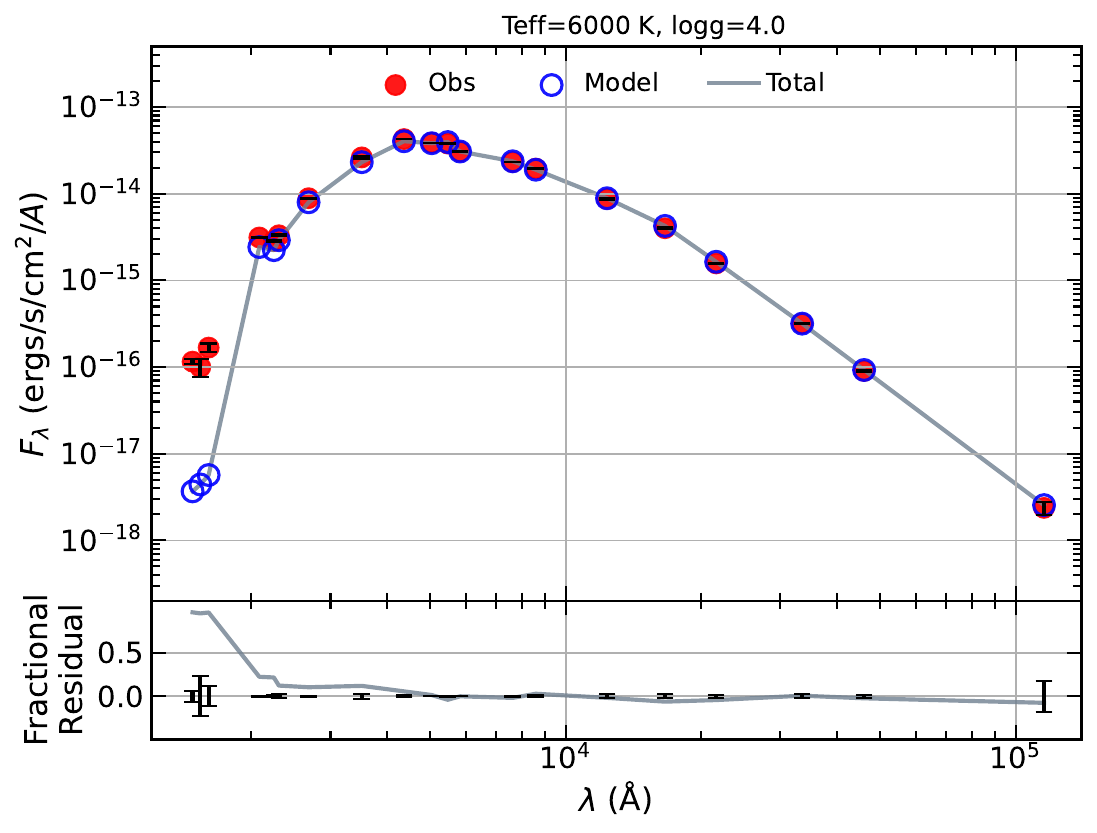}
    \caption{SED of WOCS\,6008 (left panel) fitted with a combined Kurucz (orange dash-dotted, cool) and Koester (light blue dash-dotted, hot) model; composite fit shown with a gray solid line. The right panel displays the SED of WOCS\,2003, fitted with a single model. The fractional residual plots in the lower panels show the quality of the final fits and the excess in the UV regime relative to the single-model fit. Best-fit parameters are shown on top of each plot.}
    \label{sed_fit}
\end{figure*}

 Following the conclusion that the UV flux of WOCS\,6008 likely comes from a hot companion, we employed a binary SED fitting to derive its properties. We employed the VOSA binary-fitting tool to fit the hotter part of the SED. The VOSA binary-fitting tool simultaneously fits the hotter and cooler components of a binary system by assuming that the observed flux is a linear combination of two different models. Based on spectroscopic monitoring studies, the companion of WOCS\,6008 is expected to be compact in nature \citep{1990AJ....100.1859M}. Therefore, we utilized the Koester WD model to fit the hotter component \citep{2009ApJ...696.1755T, 2010MmSAI..81..921K}.
 The Koester model grid spans a range of atmospheric parameters, with T$_{\mathrm{eff}}$ between 5,000 and 80,000~K, and $\log g$ from 6.5 to 9.5~dex. We adopted a range of $\log g$ from 6.5--9.0~dex for our fits \citep{Crumpler2025}, although the broadband fluxes used in the SED are insensitive to this choice. Overall, the combined model flux reproduces the observed SED of WOCS\,6008 well as demonstrated in Figure~\ref{sed_fit} (left panel). The best-fit parameters of WOCS\,6008, derived from the best SED fit with the lowest values of $\chi^2_{red}$ and Vgf$_{b}$, are listed in Table~\ref{tab3}. We obtained its Vgf$_{b}$ values below 15, indicating a good SED fit and confirming the reliability of all derived stellar parameters. Further, to assess the evolutionary stage and the nature of the hot companion, we place it in the HR diagram as discussed in the subsequent section.

\begin{table*}
  	\begin{center}
	\caption{Best-fitting stellar parameters of cool (A) and hot (B) components of WOCS\,6008 presented in this paper. The first column lists the WOCS\,ID number,
    and the second column identifies the component of the system. The different models used to fit the cooler and hotter components of the SEDs are presented in the third column. The T$_{\mathrm{eff}}$, Luminosity, and Radius estimated for Li-rich star \& WD companion, along with errors, are tabulated in columns 4th, 5th, and 6th, respectively. Columns 7th, 8th, and 9th list the reduced $\chi^{2}$, vgf and $Vgf_{b}$ values corresponding to the best fit. The ratio of the number of photometric data points ($\frac{N_{fit}}{N_{tot}}$) used for the fit to the total number of available data points is listed in the 10th column.}
	\label{tab3}
    \small
   \resizebox{1.89\columnwidth}{!}{%
\begin{tabular}{ccccccccc} 
		\hline
		\hline
		 WOCS\,ID & Type &  Model Used & {\it T$_{\mathrm{eff}}$} (K) & $\frac{L}{L_{\rm{\odot}}}$  & $\frac{R}{R_{\rm{\odot}}}$ & ${\chi}_{red}^2$ & Vgf & $Vgf_{b}$ \\
         
		\hline
   
		6008 & A & Kurucz & $5,750 \pm 125$ & $5.99 \pm 0.25$ & $2.47 \pm 0.051$ & 10.4 & 2.57 & 0.14 \\
		 & B & Koester & $9,250 \pm 125$ & $0.028 \pm 0.002$ & $0.066 \pm 0.001$ & 14.3 & 6.0 & 0.59 \\
		\hline 
	\end{tabular}
  }
	\end{center}
\end{table*}

\section{The Hot Companion of WOCS\,6008}
\label{sec:hot_comp}
\begin{figure}
    \centering
    \includegraphics[width=\columnwidth]{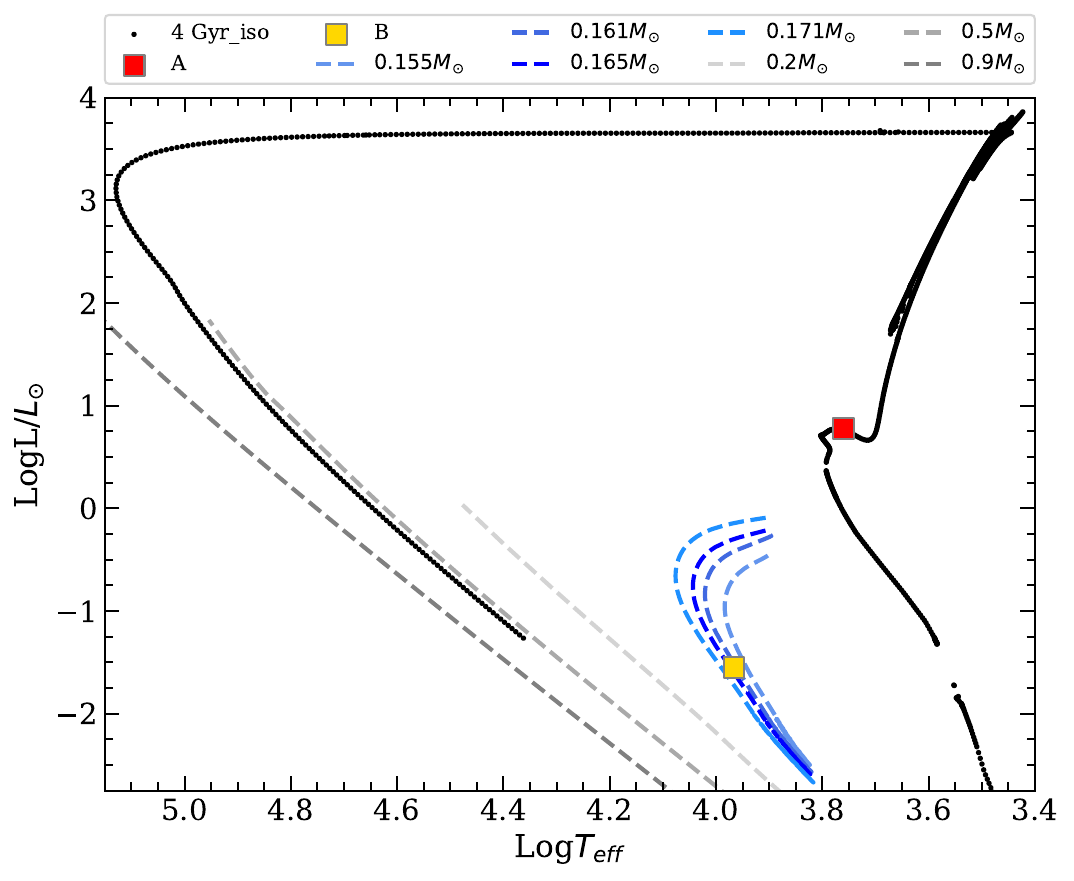}
    \caption{HR diagram of M67 showing the location of two components of Li-rich WOCS\,6008 binary system. The hot component is depicted with a yellow square symbol, whereas the cool component is displayed with a red square symbol. The dashed blue and gray lines correspond to ELM WD \citep{2013A&A...557A..19A} and WD \citep{2011ApJ...730..128T} cooling sequences with different masses, respectively. The MIST evolutionary track with solar metallicity and cluster age is also shown with black dots.}
    \label{fig:hrd}
\end{figure}

In order to obtain the masses and ages of hot companions, we employed stellar models, which provide more precise but model-dependent determinations of both parameters. We constructed a Hertzsprung-Russell (HR) diagram using stellar parameters such as T$_{\mathrm{eff}}$ and L derived from SED modeling to assess the evolutionary status of both components in the WOCS\,6008 system. It can be seen from the estimated radius of the hot companion that it is compact in nature and probably belongs to the WD class. To probe its nature, we first compared its T$_{\mathrm{eff}}$ and luminosity with the DA-type helium (He) and carbon-oxygen (CO) WD cooling sequences, with masses ranging from 0.2 to 1.3 $M_{\odot}$, provided by \cite{2011ApJ...730..128T}. However, our object’s position did not align with these tracks, shown as gray dashed lines in Figure~\ref {fig:hrd}. We then compared its location with theoretical evolutionary sequences computed for He ELM WDs and plotted them in the HR diagram. \cite{2013A&A...557A..19A} provided evolutionary sequences for He core WDs across their expected mass range ($0.15\leq M_{*}/M_{\odot}\leq 0.45$), including extremely low mass (ELM) WDs \citep[$M_{*}/M_{\odot}<0.3$,][]{2013ApJ...769...66B}. Here, we present only those evolutionary sequences of ELM WDs with masses of 0.155$M_{\odot}$, 0.161$M_{\odot}$, 0.165$M_{\odot}$, and 0.171$M_{\odot}$ that match the position of the hot companion in the HR diagram as shown in Figure~\ref {fig:hrd}. For the cool component of this system, we plotted the MIST theoretical evolutionary sequences from MS to the tip of the RGB stage from \cite{2016ApJ...823..102C}. Figure~\ref{fig:hrd} illustrates the HR diagram with its cool and hot components marked in red and yellow, respectively, and the ELM WD tracks as blue dashed lines. 

In the aforementioned models, the cooling age of the WD is estimated from the point marking the end of mass transfer and WD formation. The age and mass of a given star are determined based on the minimum distance between the star and the evolutionary track in the HR diagram. We have listed the age and mass of the hot companion, along with the model used to determine them, in Table~\ref{tab4}. For the two components of WOCS\,6008, which we refer to as A and B, we derive radii and luminosities of $R_{A}$ = 2.47$\pm$0.05 $R_{\odot}$, $L_{A}$ = 5.99$\pm$0.25 $L_{\odot}$, and $R_{B}$ = 0.07$\pm$0.001 $R_{\odot}$, $L_{B}$ = 0.028$\pm$0.002 $L_{\odot}$, respectively. These values are consistent with those expected for a subgiant star and an ELM WD.

We conclude that the most likely companion is, in fact, an ELM WD with a mass of $M_{B} = 0.161\pm0.01 M_{\odot}$, and an age of 2.5 Gyr. Furthermore, we calculated the mass of the cool companion of WOCS\,6008 by comparing it with predictions from stellar evolutionary calculations. Note that these models assume single star evolution and do not account for binary interaction effects; detailed binary evolution modeling to constrain the mass of cool companion is beyond the scope of this paper. We tried different stellar isochrones, including BaSTI \citep{2018ApJ...856..125H}, PARSEC \citep{2012MNRAS.427..127B}, and MIST \citep{2016ApJ...823..102C}, and obtained consistent results. We therefore infer the mass of the subgiant using MIST isochrones to be $M_{A}$ = $1.32\pm0.03 M_{\odot}$, using the cluster age and metallicity listed in Table~\ref{tab4} and the reddening discussed in Sect.~\ref{sec:data}. The measured mass of the ELM WD is consistent with its kinematic mass ($0.15\pm0.01M_{\odot}$, lower limit), estimated using the available mass function, and the mass of the accretor \citep{1990AJ....100.1859M}. 

Mass transfer in binaries, depending on the evolutionary stage of the donor star, is classified as Case A, Case B, or Case C \citep{Paczynski1971}. For instance, if the donor is in the MS phase, then it indicates Case A mass transfer. If mass transfer occurs after the exhaustion of core hydrogen burning but before the onset of core helium burning, it is designated as Case B. If it occurs after core helium exhaustion, typically during the AGB phase, then it corresponds to Case C. In the case of low mass stars, after hydrogen exhaustion, the mass of the He core is approximately 0.1 $\rm M_{\odot}$ or less \citep{Kippenhahnetal2012, 2016ApJ...823..102C}. When a star evolves off the MS turn-off to the giant phase, its core mass increases, and it ranges from 0.1-0.45 $\rm M_{\odot}$. If mass transfer occurred through early Case B before ascending the RGB, then the donor star will end up as a WD with a mass of $\sim0.2 \rm M_{\odot}$ or less, while a WD with a mass greater than $\sim0.45 \rm M_{\odot}$ is predicted in the case of AGB mass transfer. The derived mass of the hot companion ($0.161 \rm M_{\odot}$) in WOCS\,6008 suggests early case B mass transfer. Modeling this system could further help constrain its mass transfer history and formation, but we could not perform modeling because all available models to date simulate circular binaries, not eccentric binaries.

\begin{table*}
    \begin{center}
    \large
     \caption{Kinematic details of Li-rich star listed along with its age and mass of the WD companion. The first column presents the WOCS ID, followed by the stellar mass function, and the model used to obtain the age and mass of the companion. The last two columns indicate the lower limit of the mass of the companion and the mass of the Li-rich star.}
    \label{tab4}
      \resizebox{0.79\linewidth}{!}{%
	\begin{tabular}{cccccccc} 
		\hline
        \hline
		 WOCS\,ID & f(m) & Model used & WD age & WD mass & WD kinematic mass & M$_{Li-rich}$ & \\		
   & (10$^{-3}$M$_{\odot}$) & & (Myr) & (M$_{\odot}$) & (M$_{\odot}$) & (M$_{\odot}$) & \\
  \hline
         6008 & $1.6 \pm 0.13$ & Althaus+2013 & 2528 & $0.161 \pm 0.01$ & $0.152\pm0.01$ & $1.32\pm0.03$ & \\
         \hline
	\end{tabular}
 }

 {\footnotesize \textbf{Note.} $f(m) = (M_{2}\sin i)^{3}$/$(M_{1} + M_{2})^{2}$ represents the mass function utilized for kinematic mass determination of the detected WD companion. }
\end{center}
\end{table*}

\begin{table}
  	\begin{center}
	\small
	\caption{LTE and NLTE abundances of elements derived in WOCS\,6008, along with the number of lines used to measure the abundances (N). The last two columns present the random ($\sigma_{ran}$) and systematic ($\sigma_{par}$) uncertainties, the latter arising from errors in the stellar parameters.}
	\label{tab5}
\begin{tabular}{cccccc} 
		\hline
		\hline
		 Element & N & LTE (dex) & NLTE (dex) & $\sigma_{ran}$ (dex) & $\sigma_{par}$ (dex)  \\
		\hline
		$A(\mathrm{Li})$ & 1 & +2.67 & +2.66 & 0.01 & 0.12\\ 
        {[C/Fe]} & 3 & -0.02 & -- & 0.01 & 0.08\\
        {[Na/Fe]} & 5 & 0.0 & -0.12 & 0.02 & 0.08\\
        {[Mg/Fe]} & 5 & 0.13 & 0.14 & 0.04 & 0.07\\
        {[Al/Fe]} & 2 & -0.04 & -- & 0.04 & 0.06\\
        {[Y/Fe]} & 10 & 0.02 & 0.04 & 0.06 & 0.06\\
        {[Zr/Fe]} & 2 & 0.26 & -- & 0.09 & 0.16\\
        {[Ba/Fe]} & 2 & 0.33 & 0.27 & 0.01 & 0.08\\
        {[La/Fe]} & 5 & -0.06 & -- & 0.05 & 0.09\\
        {[Ce/Fe]} & 1 & -0.1 & -- & 0.06 & 0.06\\
		\hline 
	\end{tabular}
	\end{center}
\end{table}

\begin{figure*}
    \centering
    \includegraphics[width=2.0\columnwidth]{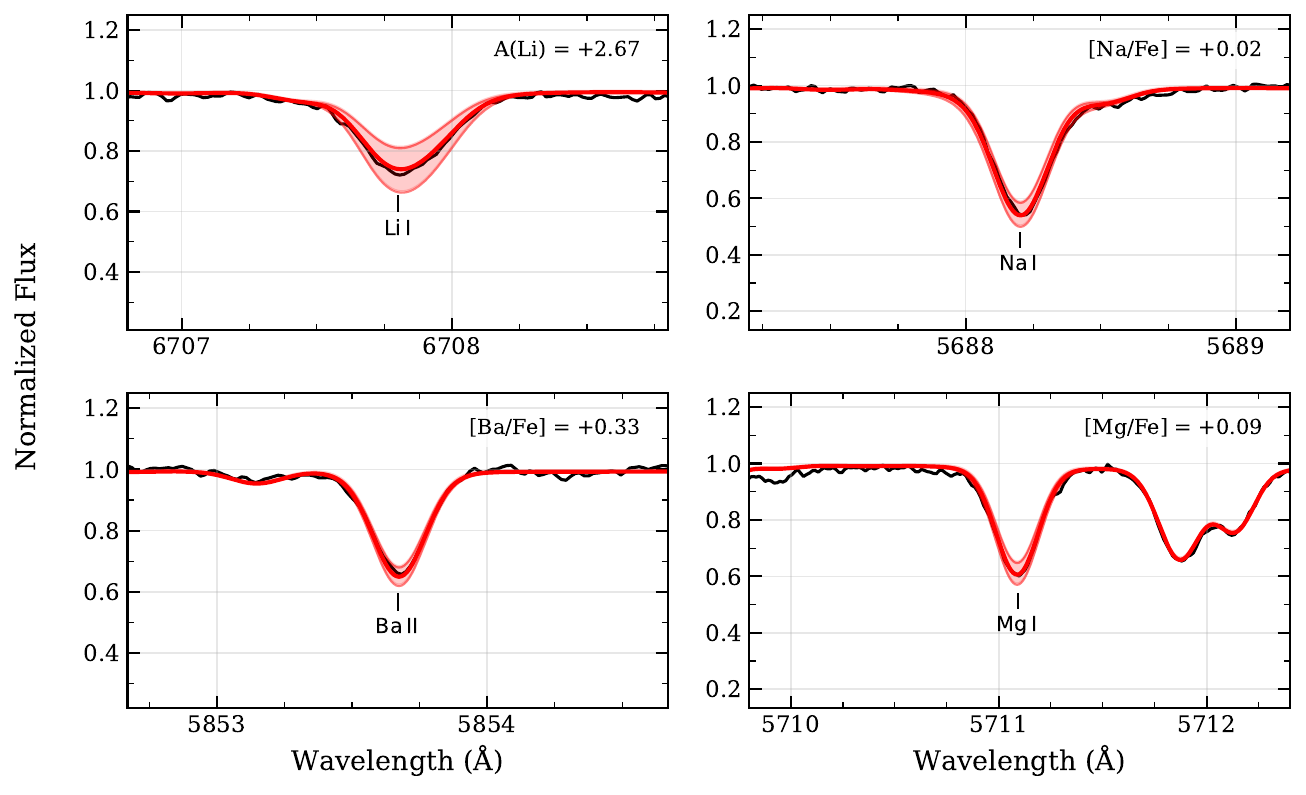}
    \caption{Spectral regions of WOCS\,6008 spectra around the Li\,I, Na\,I, Ba\,II, and Mg\,I lines displayed in the upper and lower panels. In each panel, the observed spectra are shown as a solid black line, and the best-fitting synthetic spectra are shown as a red line. The red-shaded region represents the synthetic spectra with an abundance variation of $\pm$0.2 with respect to the best fitted one.}
    \label{fig:spec}
\end{figure*}

\section{Photometric Variability}
\label{sec:variable}
There was some confusion in the literature about the photometric binary nature of WOCS\,6008, and we conducted a deeper search to investigate the matter. \cite{2015A&A...573A.100F} evaluated the efficiency of their variable star detection methodology, they determined whether any variable stars in the Wide-Field near-IR CAMera (WFCAM) calibration fields that are known from other catalogs were not detected by their method. By cross-matching their initial sample of 216,722 light curves with all survey catalogs of variable stars included in the International Virtual Observatory Alliance (IVOA), they found that 15 stars were missed due to the low amplitude of their variability. Therefore, they flagged those sources as non-detected variables in their catalog, and showed the light curves of several stars as examples of non-detected variables in their Figure 10, including WOCS\,6008 (WVSC$-$333). The light curves of WOCS\,6008, shown in the upper right panel of Figure 10 in \cite{2015A&A...573A.100F}, exhibit no variability, as noted by the authors. In fact, we checked the previous existing catalogs of variable stars \citep{2006SASS...25...47W} in which it is classified as an uncertain EB (OID 56500). In the catalog of Suspected Variable stars supplement \citep[ID: NSV 18065,][]{1998IBVS.4655....1K}, it is not identified as a variable star at all. 

\cite{2023A&A...672A.159G} analyzed Kepler K2\footnote{To double-check, we also analyzed the Kepler light curves using data from campaigns 5, 16, and 18 with the Lomb-Scargle method. We did not find any signature of an EB; rather, all of them showed a strong and consistent short-period signal around 6.5 days. We therefore confirm that it is not an EB.} light curves of M67 stars to investigate the overall trend of rotation within the cluster, including our target (EPIC ID: 211408858). Indeed, this star shows photometric variability, but the source of variability is rotation with a period of 6.54 days. They flagged it as a non-photometric binary in their catalog. Therefore, an earlier tentative classification as an EB is not supported by other datasets.

\section{Abundance Analysis}
\label{sec:chem}
The determination of the stellar atmospheric parameters, i.e., T$_{\mathrm{eff}}$, log\,{\it g}, and microturbulent velocity $\rm v_{t}$, is essential for obtaining elemental abundances. The T$_{\mathrm{eff}}$ was estimated using the photometric (B-V) color. We adopted photometric B and V magnitudes for this star from \cite{1993AJ....106..181M}, and a reddening value of E(B$-$V) = 0.04\,mag \citep{2021MNRAS.504..356D}. The metallicity of [Fe/H] = -0.05~dex was taken from \cite{2006A&A...451..993C}. We utilized the calibrated B$-$V color – T$_{\mathrm{eff}}$ transformation, especially valid for dwarfs and subgiants by \cite{2010A&A...512A..54C}. We derived T$_{\mathrm{eff}}$ = 5770 $\pm$ 130\,K, which is consistent with previous literature estimates \citep{2006A&A...451..993C} and with SED-based estimate in this paper. The uncertainty in T$_{\mathrm{eff}}$ includes those arising from the photometric color, the metallicity, and the adopted color – T$_{\mathrm{eff}}$ transformation.
The log\,{\it g} is determined using the following relation \citep{2024A&A...689A..89M}:
\begin{equation}
    log\,{\it g} = -10.32 + log(M) + 4.T_{\mathrm{eff}} - log(L)
\end{equation}

To calculate the stellar luminosity (log($L/L_{\odot}$ = -0.4\,.\,($M_{bol} - M_{\odot}$), we adopted the distance to the cluster 865 $\pm$ 18 pc, and the bolometric correction estimated using the tool available at Gaia DR3 website\footnote{\url{https://www.cosmos.esa.int/web/gaia/dr3-bolometric-correction-tool}} \citep{2023A&A...674A..26C}.

We employed the MIST isochrone generated for the cluster parameters to estimate the mass of the star, as listed in Table~\ref{tab4}. We calculated the log\,{\it g} value of 4.0$\pm$0.1, where the uncertainty accounts for uncertainties in the T$_{\mathrm{eff}}$, luminosity, and adopted stellar mass. The microturbulence, $\rm v_{t}$, was estimated using the relation calibrated for dwarfs and subgiants provided by \cite{2013A&A...555A.150T}. The final atmospheric parameters adopted for WOCS\,6008 to compute synthetic spectra are $\rm 5570\pm130~K$ for $\rm T_{eff}$, $\rm 4.0\pm0.1~dex$ for log\,{\it g}, $\rm -0.05\pm0.1~dex$ for [Fe/H], and $\rm 1.1\pm0.1~km s^{-1}$ for $\rm v_{t}$.

Abundances of Li, C, Na, Mg, Al, Fe, and s-process elements were derived through spectral synthesis. To carry out the analysis, we utilized the proprietary code SALVADOR (Alvarez Garay et al., in prep.), which determines abundances through $\chi^{2}$ minimization between the observed spectral line and a grid of synthetic spectra. The synthetic spectra were computed on the fly with the SYNTHE code \citep{2004MSAIS...5...93S, 2005MSAIS...8...14K}, varying solely the abundance of the element under analysis. The synthetic spectra were calculated by adopting one-dimensional, plane-parallel, local thermodynamic equilibrium (LTE) model atmospheres employing the ATLAS9 code \citep{2004MSAIS...5...93S, 2005MSAIS...8...14K} and the new opacity distribution functions of the KOALA database \citep{2026A&A...705A.134M}.

For all the elements considered for analysis, we selected unblended and unsaturated neutral and singly ionized lines, as listed in the Table~\ref{tab6} provided in Appendix~\ref{app:append}. Since continuum normalization is critical for obtaining reliable abundances, we have performed local continuum normalization within a 10 $\AA$ region around the line to minimize uncertainties in the resulting abundances. The total uncertainties on the abundance determination include both the internal error and the systematic error, primarily associated with the choice of stellar atmospheric parameters. The internal error on the abundance of an element X is estimated as $\sigma /\sqrt{N}$, where $\sigma$ is the root mean square (rms) around the mean X abundance, and N is the number of lines of the X element (with N $\geq$ 2). In the case of elements with only one transition (N=1), it will estimate uncertainty using Monte Carlo simulations. Systematic uncertainties are dominated by the errors in the choice of stellar parameters. They were estimated by varying T$_{\mathrm{eff}}$ by 130 K, log\,{\it g} by 0.1~dex and $v_{t}$ by 0.1 km s$^{−1}$. Figure~\ref{fig:spec} shows the best-fit synthetic spectrum to the observed spectrum around the Li\,I, Na\,I, Ba\,II, and Mg\,I lines, where the red-shaded region marks the synthetic spectra with an abundance variation of $\pm$0.2~dex relative to the best-fit.

We remeasured the Fe abundance for this star using a suitable linelist of unblended and unsaturated lines, and derived [Fe/H] = $-0.01\pm0.01$~dex, which is in excellent agreement with values reported in the literature. As a reference, solar abundances are adopted from \cite{2010ASSP...16..379L} and \cite{2011SoPh..268..255C}. The Li abundance ($A(\mathrm{Li}) = 2.67~\mathrm{dex}$) was also derived from the resonance line at 6708 \AA, and is consistent with its previous estimate \citep{2006A&A...451..993C}. We have taken into account the corrections for the departure from local thermal equilibrium (LTE) for those elements for which non-LTE (NLTE) corrections are available in the literature. Here we report both LTE and NLTE abundances, in the latter case, corrections were adopted from \cite{2011A&A...528A.103L} for Li and Na, from \cite{2017ApJ...847...15B} for Mg, from \cite{2023ApJ...957...10A} for Y. We corrected the Ba abundances derived from the lines at 5853 and 6141 Å for the NLTE effects by interpolating them into the grids of \cite{2015A&A...581A..70K}. The LTE and NLTE abundances of all the considered elements in this work are listed in Table~\ref{tab5}. \cite{2023A&A...676A.129H} estimated stellar parameters and elemental abundances using the \textit{Gaia} ESO spectroscopic data for more than 114000 stars, including WOCS\,6008. Our derived abundances of all elements are consistent with their estimates.

The abundances of all s-process elements are found to be solar within uncertainties, with the exception of Ba, which is slightly enhanced with a [Ba/Fe]$_{\mathrm{LTE}}$ value of 0.33$\pm$0.01~dex. C ($-0.02\pm0.01$), Na ($-0.12\pm0.02$) and Al ($-0.04\pm0.04$) abundances are solar, whereas Mg ($0.14\pm0.04$) is slightly enhanced. If mass transfer had occurred during the AGB phase, it would have resulted in enhanced Na and depleted Mg. Another chemical signature indicative of mass transfer during the AGB phase is the enrichment of s-process elements. It should be noted that the formation of s-process elements occurs during the thermally pulsing phase of low- and intermediate-mass AGB stars, which in turn explains the s-process enhancement typically found in chemically peculiar stars \citep{1999ARA&A..37..239B, 2004ApJ...601..864T, 2014PASA...31...30K}. However, in case mass transfer happens before the thermal pulsing phase in AGB stars, no s-process enhancement is expected. We did not find any signature of chemical peculiarity, which suggests that mass transfer during the AGB phase is unlikely. Since stars undergoing the first dredge-up on the RGB exhibit a significantly lowered envelope C/N ratio \citep{1967ApJ...147..624I}, an altered C/N ratio in our target star would indicate past mass transfer from an RGB companion. To test this, we derived the C abundance using isolated atomic lines in our spectra, and found it to be solar within the uncertainties. Nevertheless, we could not obtain an abundance of N because the only available ${}^{12}\mathrm{C}{}^{14}\mathrm{N}$ red system bands in the region $6470-6490\text{ \AA}$ were not detected in our spectra \citep{ 2025A&A...703A...4T}. While molecular bands typically appear in cool stars with temperatures around $4000\text{ K}$ \citep{1973A&A....23..411T, 2018PASA...35...46B}, our star's temperature ($5800\text{ K}$) is too high to show molecular features. On the basis of the estimate of the C abundance ratio, it might be possible that the N abundance is also solar rather than enhanced by the first dredge-up \citep{2019ApJ...874...97S}. If this is the case, mass transfer during the RGB phase is unlikely. We therefore conclude that mass transfer in WOCS\,6008 did likely not occur through Case C or late Case B, but rather through early Case B. This is supported by the derived mass of the hot companion, i.e., $0.161\pm0.01 \rm M_{\odot}$, which is also indicative of early Case B mass transfer, as discussed in Sect.~\ref{sec:hot_comp}. 

\section{Discussion}
\label{sec:dis}
The origin of Li-rich stars remains an open question. In this section, we discuss potential scenarios that explain the observed Li excess in the two known Li-rich stars, WOCS\,2003 and WOCS\,6008. As WOCS\,2003 is a short-period tidally locked binary (SB2), \cite{1994ApJ...434L..71D} proposed a combination of Zahn’s tidal circularization and rotational mixing to explain its Li enrichment. In this study, we did not constrain its formation pathways because its well-known high activity levels fully explain the UV excess, and the use of SED binary-fitting in this case would not be useful. We found one archival spectrum of WOCS\,2003, but since it is an SB2, a reliable abundance analysis would require a spectral disentangling technique, which, in turn, requires 10-15 epochs to provide accurate results \citep{2026A&A...706A.179A}. Moreover, since both stars are on the MS, we do not expect to see any particular chemical signatures that would shed additional light on the system's nature.

\par
Considering that WOCS\,6008 is indeed part of a binary system, we focus our detailed discussion on its formation. We first discuss the less likely scenarios proposed to explain the formation of Li-rich stars, as described in Sect.~\ref{sec:intro}, followed by the most favored pathways. Since this star is in the SGB evolutionary phase, the Li excess cannot be explained by intrinsic nucleosynthesis and post-first dredge-up mixing. While the CF mechanism produces Li via hot bottom burning, this process is triggered only in AGB stars with masses greater than $4 \rm M_{\odot}$, or in RGB stars undergoing extra mixing; it is not active in MS stars. Hence, pollution from an external source, such as a highly evolved RGB or AGB companion, could explain its Li excess. However, based on our mass estimate of the hot companion and the measured abundances of s-process and other light elements, we rule out Li enhancement via mass transfer from an AGB or RGB companion. The scenario involving the engulfment of planets or planetesimal material \citep{1999MNRAS.308.1133S} appears unlikely for this system, as the radius of the star at the SGB stage is not sufficient to engulf such substellar systems. Another scenario, proposed by \cite{1978ApJ...222..600S}, suggests that Li is produced in novae via thermonuclear runaway. This material can subsequently pollute the interstellar medium and account for the stellar Li enrichment observed in the Milky Way \citep{2015ApJ...808L..14I}. A chemical signature favoring this mechanism is the detection of high beryllium (Be) enrichment, which is expected during a nova outburst and has been confirmed in previous studies \citep{2016MNRAS.463L.117M, 2020MNRAS.492.4975M, 2022MNRAS.509.3258M}. \cite{2020ApJ...888...28B} derived Be abundances for nine subgiants in M67 for which Li abundances had been previously determined. They showed that Be abundances also decline across the SGB, though less steeply than Li. Our target was not included in their sample; therefore, no Be abundance estimate is available. Moreover, the optical spectra analyzed in this work do not cover the 3100 $\AA$ region required to detect the Be resonance feature. However, the WDs identified in classical and recurrent novae systems consistently span a mass range of $\sim 0.5 - 1.4 \rm M_\odot$ \citep[e.g.,][]{1991ApJ...376..177R, 2010ApJS..187..275S}. Our detected He WD has a low mass of $\sim 0.2 \rm M_\odot$; such low-mass WDs lack both the high surface gravity and the carbon-oxygen catalyst necessary to trigger a thermonuclear runaway. Thus, a nova explanation also seems highly unlikely.

\par
The most viable scenario to explain the Li enrichment in WOCS\,6008 is tidal spin-up by a binary companion, where, as proposed by \cite{1994A&A...288..829Z}, tidal effects acting at the periastron stage of WOCS\,6008 lead the system to pseudo-synchronization and therefore produce a spin-up. This results in enhanced coronal and chromospheric activities observed in this system. The presence of this spin-up is also reinforced by the rotational velocity estimated for this system. \cite{2023A&A...672A.159G} estimated the rotation periods of FGK type stars in M67 using the Kepler K2 data, including WOCS\,6008, and found its rotation period of $6.54\pm0.1$ days. Using the radius of 2.47R$_{\odot}$ derived in this work and the known rotation period of 6.54 days, we estimate its average rotational velocity of about 19 km\,s$^{-1}$ and corresponding inclination angle of 18$^\circ$ (vsini = 6 km\,s$^{-1}$). This velocity is at least three times higher than expected for single subgiants or unsynchronized binaries with subgiant components \citep{1996A&A...314..499D}. Thus, its high rotational velocity further confirms the spin-up \citep{2006A&A...451..993C}. If this spin-up is efficient enough in strength and duration, it aids in preserving the Li present at the previous turn-off stage \citep{Jeffries2020, Andres2021}. Moreover, as shown in Fig.~2 of \cite{2012A&A...541A.150P}, there are five MS stars with Li abundances similar to that of WOCS~6008. This indicates that Li is preserved in this star through tidal synchronization, as discussed above. Alternatively, \cite{2006A&A...451..993C} suggested atomic diffusion as another route that can account for the high Li abundance observed at this evolved stage. The high rotational velocity increases the diffusion coefficient ($D_{mix}$), which is proportional to the square of the rotational velocity \citep{2019ApJ...880..125C}. According to models that include diffusion and rotation-induced mixing, a significant increase in the $D_{mix}$ value can account for the observed Li abundance in the photosphere \citep{2003ApJ...593..509D}. Another possible mechanism is the accretion of Li-rich material from the donor star, prior to the canonical mixing that triggers Li depletion in later stages. From the above considerations, we infer that rotation and/or atomic diffusion are the most favored mechanisms that account for Li enrichment in this post-mass-transfer system.

\section{Summary and Conclusions}
\label{sec:summary}
This is the first study to trace possible hot companions to Li-rich stars in the old OC M67. We compiled archival UV photometric data from several space-based missions, such as UVIT, {\it GALEX}, and Swift/UVOT, to characterize the two previously identified Li-rich stars in M67. Furthermore, we employed archival high-resolution UVES spectra to find the chemical signatures of binarity in the post-mass transfer system WOCS\,6008. In order to test the binarity hypotheses proposed to explain the enrichment of Li in Li-rich stars, as described in Sect.~\ref{sec:intro}, we utilized a multiwavelength SED fitting technique to detect and characterize the potential hot companions, if any. We found that two known Li-rich stars exhibit significant excess flux in the FUV, though the underlying causes are entirely different: chromospheric and coronal activity accounts for the excess flux in WOCS\,2003, whereas a hot companion is responsible in WOCS\,6008.\par 

Thus, we report the first detection of an unseen companion to WOCS\,6008. By comparing its location in the HR diagram with theoretical evolutionary sequences for ELM WDs, we estimate the companion's mass to be $0.161 \pm 0.01 \rm M_\odot$, which is consistent with its kinematic mass. Based on its mass, we suggest that the companion is a helium core ELM WD with a cooling age of 2.5 Gyr, indicating that mass transfer occurred via early Case B. 
Determination of s-process and light elements abundances in WOCS\,6008 revealed no enhancement or depletion, indicating that mass transfer likely did not occur in the AGB or RGB phase, in agreement with the estimated mass of the hot companion.\par

We would like to emphasize that UV excess alone cannot be attributed to the presence of a hot companion, as stellar activity can contribute significantly to the UV flux. Additional information, like a star's orbital period, mass, and activity indicators, is always required to break the degeneracy and to better characterize a given system. Based on the analysis, our study suggests rotation or atomic diffusion as a possible explanation for Li enrichment in WOCS\,6008.\par 
In the future, even low-resolution UV spectra of such systems, like those from Mauve \citep{2026RASTI...5ag018M}, will allow us to verify the presence and nature of the hot companions and disentangle their signatures from stellar activity. Moreover, modeling of these systems will further elucidate their formation mechanisms. A systematic investigation of them in diverse environments, including accurate modeling, is sorely needed to firmly assess the formation channels of Li-rich stars and their relative importance.   

\begin{acknowledgements}
      Funded by the European Union (ERC-2022-AdG, {\em "StarDance: the non-canonical evolution of stars in clusters"}, Grant Agreement 101093572, PI: E. Pancino). Views and opinions expressed are, however, those of the author(s) only and do not necessarily reflect those of the European Union or the European Research Council. Neither the European Union nor the granting authority can be held responsible for them. This publication utilizes data from the \textit{AstroSat} mission’s UVIT, archived at the Indian Space Science Data Centre (ISSDC). The UVIT project is a collaboration between IIA Bengaluru, IUCAA Pune, TIFR Mumbai, several ISRO centres, and CSA. This research also made use of VOSA, developed under the Spanish Virtual Observatory project, supported by the Spanish MINECO through grant AyA2017-84089. EL acknowledges support from the ERC Consolidator Grant funding scheme (project ASTEROCHRONOMETRY, G.A. n. 772293)
\end{acknowledgements}

\textbf{Tools}: This work made use of the following Python packages: Matplotlib \citep{4160265}, NumPy \citep{5725236}, Scipy \citep{4160250, 5725235}, Astropy \citep{refId0, 2018AJ....156..123A}, and Pandas \citep{mckinney-proc-scipy-2010}. Data exploration relied on Topcat \citep{2011ascl.soft01010T}.

\bibliographystyle{aa}
\bibliography{references}

@ARTICLE{2017AJ....154..128T,
       author = {{Tandon}, S.~N. and {Subramaniam}, Annapurni and {Girish}, V. and {Postma}, J. and {Sankarasubramanian}, K. and {Sriram}, S. and {Stalin}, C.~S. and {Mondal}, C. and {Sahu}, S. and {Joseph}, P. and {Hutchings}, J. and {Ghosh}, S.~K. and {Barve}, I.~V. and {George}, K. and {Kamath}, P.~U. and {Kathiravan}, S. and {Kumar}, A. and {Lancelot}, J.~P. and {Leahy}, D. and {Mahesh}, P.~K. and {Mohan}, R. and {Nagabhushana}, S. and {Pati}, A.~K. and {Kameswara Rao}, N. and {Sreedhar}, Y.~H. and {Sreekumar}, P.},
        title = "{In-orbit Calibrations of the Ultraviolet Imaging Telescope}",
      journal = {\aj},
         year = 2017,
        month = sep,
       volume = {154},
       number = {3},
          eid = {128},
        pages = {128},
          doi = {10.3847/1538-3881/aa8451},
archivePrefix = {arXiv},
       eprint = {1705.03715},
 primaryClass = {astro-ph.IM},
       adsurl = {https://ui.adsabs.harvard.edu/abs/2017AJ....154..128T}
}

@ARTICLE{2020AJ....159..158T,
       author = {{Tandon}, S.~N. and {Postma}, J. and {Joseph}, P. and {Devaraj}, A. and {Subramaniam}, A. and {Barve}, I.~V. and {George}, K. and {Ghosh}, S.~K. and {Girish}, V. and {Hutchings}, J.~B. and {Kamath}, P.~U. and {Kathiravan}, S. and {Kumar}, A. and {Lancelot}, J.~P. and {Leahy}, D. and {Mahesh}, P.~K. and {Mohan}, R. and {Nagabhushana}, S. and {Pati}, A.~K. and {Rao}, N. Kameswara and {Sankarasubramanian}, K. and {Sriram}, S. and {Stalin}, C.~S.},
        title = "{Additional Calibration of the Ultraviolet Imaging Telescope on Board AstroSat}",
      journal = {\aj},
         year = 2020,
        month = apr,
       volume = {159},
       number = {4},
          eid = {158},
        pages = {158},
          doi = {10.3847/1538-3881/ab72a3},
archivePrefix = {arXiv},
       eprint = {2002.01159},
 primaryClass = {astro-ph.IM},
       adsurl = {https://ui.adsabs.harvard.edu/abs/2020AJ....159..158T}
}

@ARTICLE{1987PASP...99..191S,
       author = {{Stetson}, Peter B.},
        title = "{DAOPHOT: A Computer Program for Crowded-Field Stellar Photometry}",
      journal = {\pasp},
         year = 1987,
        month = mar,
       volume = {99},
        pages = {191},
          doi = {10.1086/131977},
       adsurl = {https://ui.adsabs.harvard.edu/abs/1987PASP...99..191S}
}

@ARTICLE{2021MNRAS.504..356D,
       author = {{Dias}, W.~S. and {Monteiro}, H. and {Moitinho}, A. and {L{\'e}pine}, J.~R.~D. and {Carraro}, G. and {Paunzen}, E. and {Alessi}, B. and {Villela}, L.},
        title = "{Updated parameters of 1743 open clusters based on Gaia DR2}",
      journal = {\mnras},
         year = 2021,
        month = jun,
       volume = {504},
       number = {1},
        pages = {356-371},
          doi = {10.1093/mnras/stab770},
archivePrefix = {arXiv},
       eprint = {2103.12829},
 primaryClass = {astro-ph.SR},
       adsurl = {https://ui.adsabs.harvard.edu/abs/2021MNRAS.504..356D}
}

@ARTICLE{2008A&A...492..277B,
       author = {{Bayo}, A. and {Rodrigo}, C. and {Barrado Y Navascu{\'e}s}, D. and {Solano}, E. and {Guti{\'e}rrez}, R. and {Morales-Calder{\'o}n}, M. and {Allard}, F.},
        title = "{VOSA: virtual observatory SED analyzer. An application to the Collinder 69 open cluster}",
      journal = {\aap},
         year = 2008,
        month = dec,
       volume = {492},
       number = {1},
        pages = {277-287},
          doi = {10.1051/0004-6361:200810395},
archivePrefix = {arXiv},
       eprint = {0808.0270},
 primaryClass = {astro-ph},
       adsurl = {https://ui.adsabs.harvard.edu/abs/2008A&A...492..277B}
}

@INPROCEEDINGS{2003IAUS..210P.A20C,
       author = {{Castelli}, F. and {Kurucz}, R.~L.},
        title = "{New Grids of ATLAS9 Model Atmospheres}",
    booktitle = {Modelling of Stellar Atmospheres},
         year = 2003,
       editor = {{Piskunov}, N. and {Weiss}, W.~W. and {Gray}, D.~F.},
       volume = {210},
       series = {IAU Symposium},
        month = jan,
        pages = {A20},
          doi = {10.48550/arXiv.astro-ph/0405087},
archivePrefix = {arXiv},
       eprint = {astro-ph/0405087},
 primaryClass = {astro-ph},
       adsurl = {https://ui.adsabs.harvard.edu/abs/2003IAUS..210P.A20C}
}

@ARTICLE{2017PASP..129k5002P,
       author = {{Postma}, Joseph E. and {Leahy}, Denis},
        title = "{CCDLAB: A Graphical User Interface FITS Image Data Reducer, Viewer, and Canadian UVIT Data Pipeline}",
      journal = {\pasp},
         year = 2017,
        month = nov,
       volume = {129},
       number = {981},
        pages = {115002},
          doi = {10.1088/1538-3873/aa8800},
       adsurl = {https://ui.adsabs.harvard.edu/abs/2017PASP..129k5002P}
}

@article{ refId0,
	author = {{Pigulski, A.} and {Kopacki, G.} and {Kolaczkowski, Z.}},
	title = {The young open cluster NGC 663 and its Be
stars*},
	DOI= "10.1051/0004-6361:20010974",
	url= "https://doi.org/10.1051/0004-6361:20010974",
	journal = {A\&A},
	year = 2001,
	volume = 376,
	number = 1,
	pages = "144-153",
}

@MISC{2011ascl.soft01010T,
       author = {{Taylor}, Mark},
        title = "{TOPCAT: Tool for OPerations on Catalogues And Tables}",
 howpublished = {Astrophysics Source Code Library, record ascl:1101.010},
         year = 2011,
        month = jan,
          eid = {ascl:1101.010},
        pages = {ascl:1101.010},
archivePrefix = {ascl},
       eprint = {1101.010},
       adsurl = {https://ui.adsabs.harvard.edu/abs/2011ascl.soft01010T}
}

@ARTICLE{4160265,
  author={Hunter, John D.},
  journal={Computing in Science \& Engineering}, 
  title={Matplotlib: A 2D Graphics Environment}, 
  year={2007},
  volume={9},
  number={3},
  pages={90-95},
  doi={10.1109/MCSE.2007.55}
}

@ARTICLE{5725236,
  author={van der Walt, Stefan and Colbert, S. Chris and Varoquaux, Gael},
  journal={Computing in Science \& Engineering}, 
  title={The NumPy Array: A Structure for Efficient Numerical Computation}, 
  year={2011},
  volume={13},
  number={2},
  pages={22-30},
  doi={10.1109/MCSE.2011.37}
}

@ARTICLE{4160250,
  author={Oliphant, Travis E.},
  journal={Computing in Science \& Engineering}, 
  title={Python for Scientific Computing}, 
  year={2007},
  volume={9},
  number={3},
  pages={10-20},
  doi={10.1109/MCSE.2007.58}
}

@ARTICLE{5725235,
  author={Millman, K. Jarrod and Aivazis, Michael},
  journal={Computing in Science \& Engineering}, 
  title={Python for Scientists and Engineers}, 
  year={2011},
  volume={13},
  number={2},
  pages={9-12},
  doi={10.1109/MCSE.2011.36}
}

@ARTICLE{2018AJ....156..123A,
       author = {{Astropy Collaboration} and {Price-Whelan}, A.~M. and {Sip{\H{o}}cz}, B.~M. and {G{\"u}nther}, H.~M. and {Lim}, P.~L. and {Crawford}, S.~M. and {Conseil}, S. and {Shupe}, D.~L. and {Craig}, M.~W. and {Dencheva}, N. and {Ginsburg}, A. and {VanderPlas}, J.~T. and {Bradley}, L.~D. and {P{\'e}rez-Su{\'a}rez}, D. and {de Val-Borro}, M. and {Aldcroft}, T.~L. and {Cruz}, K.~L. and {Robitaille}, T.~P. and {Tollerud}, E.~J. and {Ardelean}, C. and {Babej}, T. and {Bach}, Y.~P. and {Bachetti}, M. and {Bakanov}, A.~V. and {Bamford}, S.~P. and {Barentsen}, G. and {Barmby}, P. and {Baumbach}, A. and {Berry}, K.~L. and {Biscani}, F. and {Boquien}, M. and {Bostroem}, K.~A. and {Bouma}, L.~G. and {Brammer}, G.~B. and {Bray}, E.~M. and {Breytenbach}, H. and {Buddelmeijer}, H. and {Burke}, D.~J. and {Calderone}, G. and {Cano Rodr{\'\i}guez}, J.~L. and {Cara}, M. and {Cardoso}, J.~V.~M. and {Cheedella}, S. and {Copin}, Y. and {Corrales}, L. and {Crichton}, D. and {D'Avella}, D. and {Deil}, C. and {Depagne}, {\'E}. and {Dietrich}, J.~P. and {Donath}, A. and {Droettboom}, M. and {Earl}, N. and {Erben}, T. and {Fabbro}, S. and {Ferreira}, L.~A. and {Finethy}, T. and {Fox}, R.~T. and {Garrison}, L.~H. and {Gibbons}, S.~L.~J. and {Goldstein}, D.~A. and {Gommers}, R. and {Greco}, J.~P. and {Greenfield}, P. and {Groener}, A.~M. and {Grollier}, F. and {Hagen}, A. and {Hirst}, P. and {Homeier}, D. and {Horton}, A.~J. and {Hosseinzadeh}, G. and {Hu}, L. and {Hunkeler}, J.~S. and {Ivezi{\'c}}, {\v{Z}}. and {Jain}, A. and {Jenness}, T. and {Kanarek}, G. and {Kendrew}, S. and {Kern}, N.~S. and {Kerzendorf}, W.~E. and {Khvalko}, A. and {King}, J. and {Kirkby}, D. and {Kulkarni}, A.~M. and {Kumar}, A. and {Lee}, A. and {Lenz}, D. and {Littlefair}, S.~P. and {Ma}, Z. and {Macleod}, D.~M. and {Mastropietro}, M. and {McCully}, C. and {Montagnac}, S. and {Morris}, B.~M. and {Mueller}, M. and {Mumford}, S.~J. and {Muna}, D. and {Murphy}, N.~A. and {Nelson}, S. and {Nguyen}, G.~H. and {Ninan}, J.~P. and {N{\"o}the}, M. and {Ogaz}, S. and {Oh}, S. and {Parejko}, J.~K. and {Parley}, N. and {Pascual}, S. and {Patil}, R. and {Patil}, A.~A. and {Plunkett}, A.~L. and {Prochaska}, J.~X. and {Rastogi}, T. and {Reddy Janga}, V. and {Sabater}, J. and {Sakurikar}, P. and {Seifert}, M. and {Sherbert}, L.~E. and {Sherwood-Taylor}, H. and {Shih}, A.~Y. and {Sick}, J. and {Silbiger}, M.~T. and {Singanamalla}, S. and {Singer}, L.~P. and {Sladen}, P.~H. and {Sooley}, K.~A. and {Sornarajah}, S. and {Streicher}, O. and {Teuben}, P. and {Thomas}, S.~W. and {Tremblay}, G.~R. and {Turner}, J.~E.~H. and {Terr{\'o}n}, V. and {van Kerkwijk}, M.~H. and {de la Vega}, A. and {Watkins}, L.~L. and {Weaver}, B.~A. and {Whitmore}, J.~B. and {Woillez}, J. and {Zabalza}, V. and {Astropy Contributors}},
        title = "{The Astropy Project: Building an Open-science Project and Status of the v2.0 Core Package}",
      journal = {\aj},
         year = 2018,
        month = sep,
       volume = {156},
       number = {3},
          eid = {123},
        pages = {123},
          doi = {10.3847/1538-3881/aabc4f},
archivePrefix = {arXiv},
       eprint = {1801.02634},
 primaryClass = {astro-ph.IM},
       adsurl = {https://ui.adsabs.harvard.edu/abs/2018AJ....156..123A}
}

@InProceedings{ mckinney-proc-scipy-2010,
  author    = { {W}es {M}c{K}inney },
  title     = { {D}ata {S}tructures for {S}tatistical {C}omputing in {P}ython },
  booktitle = { {P}roceedings of the 9th {P}ython in {S}cience {C}onference },
  pages     = { 56 - 61 },
  year      = { 2010 },
  editor    = { {S}t\'efan van der {W}alt and {J}arrod {M}illman },
  doi       = { 10.25080/Majora-92bf1922-00a }
}

@ARTICLE{2021MNRAS.506.5201R,
       author = {{Rebassa-Mansergas}, A. and {Solano}, E. and {Jim{\'e}nez-Esteban}, F.~M. and {Torres}, S. and {Rodrigo}, C. and {Ferrer-Burjachs}, A. and {Calcaferro}, L.~M. and {Althaus}, L.~G. and {C{\'o}rsico}, A.~H.},
        title = "{White dwarf-main-sequence binaries from Gaia EDR3: the unresolved 100 pc volume-limited sample}",
      journal = {\mnras},
         year = 2021,
        month = oct,
       volume = {506},
       number = {4},
        pages = {5201-5211},
          doi = {10.1093/mnras/stab2039},
archivePrefix = {arXiv},
       eprint = {2107.06303},
 primaryClass = {astro-ph.SR},
       adsurl = {https://ui.adsabs.harvard.edu/abs/2021MNRAS.506.5201R}
}

@ARTICLE{1999PASP..111...63F,
       author = {{Fitzpatrick}, Edward L.},
        title = "{Correcting for the Effects of Interstellar Extinction}",
      journal = {\pasp},
         year = "1999",
        month = "Jan",
       volume = {111},
       number = {755},
        pages = {63-75},
          doi = {10.1086/316293},
archivePrefix = {arXiv},
       eprint = {astro-ph/9809387},
 primaryClass = {astro-ph},
       adsurl = {https://ui.adsabs.harvard.edu/abs/1999PASP..111...63F}
}

@ARTICLE{2005ApJ...619..931I,
       author = {{Indebetouw}, R. and {Mathis}, J.~S. and {Babler}, B.~L. and
         {Meade}, M.~R. and {Watson}, C. and {Whitney}, B.~A. and
         {Wolff}, M.~J. and {Wolfire}, M.~G. and {Cohen}, M. and {Bania}, T.~M. and
         {Benjamin}, R.~A. and {Clemens}, D.~P. and {Dickey}, J.~M. and
         {Jackson}, J.~M. and {Kobulnicky}, H.~A. and {Marston}, A.~P. and
         {Mercer}, E.~P. and {Stauffer}, J.~R. and {Stolovy}, S.~R. and
         {Churchwell}, E.},
        title = "{The Wavelength Dependence of Interstellar Extinction from 1.25 to 8.0 {\ensuremath{\mu}}m Using GLIMPSE Data}",
      journal = {\apj},
         year = "2005",
        month = "Feb",
       volume = {619},
       number = {2},
        pages = {931-938},
          doi = {10.1086/426679},
archivePrefix = {arXiv},
       eprint = {astro-ph/0406403},
 primaryClass = {astro-ph},
       adsurl = {https://ui.adsabs.harvard.edu/abs/2005ApJ...619..931I}
}

@ARTICLE{2011ApJ...730..128T,
       author = {{Tremblay}, P. -E. and {Bergeron}, P. and {Gianninas}, A.},
        title = "{An Improved Spectroscopic Analysis of DA White Dwarfs from the Sloan Digital Sky Survey Data Release 4}",
      journal = {\apj},
         year = 2011,
        month = apr,
       volume = {730},
       number = {2},
          eid = {128},
        pages = {128},
          doi = {10.1088/0004-637X/730/2/128},
archivePrefix = {arXiv},
       eprint = {1102.0056},
 primaryClass = {astro-ph.SR},
       adsurl = {https://ui.adsabs.harvard.edu/abs/2011ApJ...730..128T}
}

@ARTICLE{2016ApJ...823..102C,
       author = {{Choi}, Jieun and {Dotter}, Aaron and {Conroy}, Charlie and {Cantiello}, Matteo and {Paxton}, Bill and {Johnson}, Benjamin D.},
        title = "{Mesa Isochrones and Stellar Tracks (MIST). I. Solar-scaled Models}",
      journal = {\apj},
         year = 2016,
        month = jun,
       volume = {823},
       number = {2},
          eid = {102},
        pages = {102},
          doi = {10.3847/0004-637X/823/2/102},
archivePrefix = {arXiv},
       eprint = {1604.08592},
 primaryClass = {astro-ph.SR},
       adsurl = {https://ui.adsabs.harvard.edu/abs/2016ApJ...823..102C}
}

@ARTICLE{2019AJ....158...35S,
       author = {{Siegel}, Michael H. and {LaPorte}, Samuel J. and {Porterfield}, Blair L. and {Hagen}, Lea M.~Z. and {Gronwall}, Caryl A.},
        title = "{The Swift  UVOT Stars Survey. III. Photometry and Color-Magnitude Diagrams of 103 Galactic Open Clusters}",
      journal = {\aj},
         year = 2019,
        month = jul,
       volume = {158},
       number = {1},
          eid = {35},
        pages = {35},
          doi = {10.3847/1538-3881/ab21e1},
archivePrefix = {arXiv},
       eprint = {1905.04359},
 primaryClass = {astro-ph.SR},
       adsurl = {https://ui.adsabs.harvard.edu/abs/2019AJ....158...35S}
}

@ARTICLE{2023A&A...674A...1G,
       author = {{Gaia Collaboration} and {Vallenari}, A. and {Brown}, A.~G.~A. and {Prusti}, T. and {de Bruijne}, J.~H.~J. and {Arenou}, F. and {Babusiaux}, C. and {Biermann}, M. and {Creevey}, O.~L. and {Ducourant}, C. and {Evans}, D.~W. and {Eyer}, L. and {Guerra}, R. and {Hutton}, A. and {Jordi}, C. and {Klioner}, S.~A. and {Lammers}, U.~L. and {Lindegren}, L. and {Luri}, X. and {Mignard}, F. and {Panem}, C. and {Pourbaix}, D. and {Randich}, S. and {Sartoretti}, P. and {Soubiran}, C. and {Tanga}, P. and {Walton}, N.~A. and {Bailer-Jones}, C.~A.~L. and {Bastian}, U. and {Drimmel}, R. and {Jansen}, F. and {Katz}, D. and {Lattanzi}, M.~G. and {van Leeuwen}, F. and {Bakker}, J. and {Cacciari}, C. and {Casta{\~n}eda}, J. and {De Angeli}, F. and {Fabricius}, C. and {Fouesneau}, M. and {Fr{\'e}mat}, Y. and {Galluccio}, L. and {Guerrier}, A. and {Heiter}, U. and {Masana}, E. and {Messineo}, R. and {Mowlavi}, N. and {Nicolas}, C. and {Nienartowicz}, K. and {Pailler}, F. and {Panuzzo}, P. and {Riclet}, F. and {Roux}, W. and {Seabroke}, G.~M. and {Sordo}, R. and {Th{\'e}venin}, F. and {Gracia-Abril}, G. and {Portell}, J. and {Teyssier}, D. and {Altmann}, M. and {Andrae}, R. and {Audard}, M. and {Bellas-Velidis}, I. and {Benson}, K. and {Berthier}, J. and {Blomme}, R. and {Burgess}, P.~W. and {Busonero}, D. and {Busso}, G. and {C{\'a}novas}, H. and {Carry}, B. and {Cellino}, A. and {Cheek}, N. and {Clementini}, G. and {Damerdji}, Y. and {Davidson}, M. and {de Teodoro}, P. and {Nu{\~n}ez Campos}, M. and {Delchambre}, L. and {Dell'Oro}, A. and {Esquej}, P. and {Fern{\'a}ndez-Hern{\'a}ndez}, J. and {Fraile}, E. and {Garabato}, D. and {Garc{\'\i}a-Lario}, P. and {Gosset}, E. and {Haigron}, R. and {Halbwachs}, J. -L. and {Hambly}, N.~C. and {Harrison}, D.~L. and {Hern{\'a}ndez}, J. and {Hestroffer}, D. and {Hodgkin}, S.~T. and {Holl}, B. and {Jan{\ss}en}, K. and {Jevardat de Fombelle}, G. and {Jordan}, S. and {Krone-Martins}, A. and {Lanzafame}, A.~C. and {L{\"o}ffler}, W. and {Marchal}, O. and {Marrese}, P.~M. and {Moitinho}, A. and {Muinonen}, K. and {Osborne}, P. and {Pancino}, E. and {Pauwels}, T. and {Recio-Blanco}, A. and {Reyl{\'e}}, C. and {Riello}, M. and {Rimoldini}, L. and {Roegiers}, T. and {Rybizki}, J. and {Sarro}, L.~M. and {Siopis}, C. and {Smith}, M. and {Sozzetti}, A. and {Utrilla}, E. and {van Leeuwen}, M. and {Abbas}, U. and {{\'A}brah{\'a}m}, P. and {Abreu Aramburu}, A. and {Aerts}, C. and {Aguado}, J.~J. and {Ajaj}, M. and {Aldea-Montero}, F. and {Altavilla}, G. and {{\'A}lvarez}, M.~A. and {Alves}, J. and {Anders}, F. and {Anderson}, R.~I. and {Anglada Varela}, E. and {Antoja}, T. and {Baines}, D. and {Baker}, S.~G. and {Balaguer-N{\'u}{\~n}ez}, L. and {Balbinot}, E. and {Balog}, Z. and {Barache}, C. and {Barbato}, D. and {Barros}, M. and {Barstow}, M.~A. and {Bartolom{\'e}}, S. and {Bassilana}, J. -L. and {Bauchet}, N. and {Becciani}, U. and {Bellazzini}, M. and {Berihuete}, A. and {Bernet}, M. and {Bertone}, S. and {Bianchi}, L. and {Binnenfeld}, A. and {Blanco-Cuaresma}, S. and {Blazere}, A. and {Boch}, T. and {Bombrun}, A. and {Bossini}, D. and {Bouquillon}, S. and {Bragaglia}, A. and {Bramante}, L. and {Breedt}, E. and {Bressan}, A. and {Brouillet}, N. and {Brugaletta}, E. and {Bucciarelli}, B. and {Burlacu}, A. and {Butkevich}, A.~G. and {Buzzi}, R. and {Caffau}, E. and {Cancelliere}, R. and {Cantat-Gaudin}, T. and {Carballo}, R. and {Carlucci}, T. and {Carnerero}, M.~I. and {Carrasco}, J.~M. and {Casamiquela}, L. and {Castellani}, M. and {Castro-Ginard}, A. and {Chaoul}, L. and {Charlot}, P. and {Chemin}, L. and {Chiaramida}, V. and {Chiavassa}, A. and {Chornay}, N. and {Comoretto}, G. and {Contursi}, G. and {Cooper}, W.~J. and {Cornez}, T. and {Cowell}, S. and {Crifo}, F. and {Cropper}, M. and {Crosta}, M. and {Crowley}, C. and {Dafonte}, C. and {Dapergolas}, A. and {David}, M. and {David}, P. and {de Laverny}, P. and {De Luise}, F. and {De March}, R. and {De Ridder}, J. and {de Souza}, R. and {de Torres}, A. and {del Peloso}, E.~F. and {del Pozo}, E. and {Delbo}, M. and {Delgado}, A. and {Delisle}, J. -B. and {Demouchy}, C. and {Dharmawardena}, T.~E. and {Di Matteo}, P. and {Diakite}, S. and {Diener}, C. and {Distefano}, E. and {Dolding}, C. and {Edvardsson}, B. and {Enke}, H. and {Fabre}, C. and {Fabrizio}, M. and {Faigler}, S. and {Fedorets}, G. and {Fernique}, P. and {Fienga}, A. and {Figueras}, F. and {Fournier}, Y. and {Fouron}, C. and {Fragkoudi}, F. and {Gai}, M. and {Garcia-Gutierrez}, A. and {Garcia-Reinaldos}, M. and {Garc{\'\i}a-Torres}, M. and {Garofalo}, A. and {Gavel}, A. and {Gavras}, P. and {Gerlach}, E. and {Geyer}, R. and {Giacobbe}, P. and {Gilmore}, G. and {Girona}, S. and {Giuffrida}, G. and {Gomel}, R. and {Gomez}, A. and {Gonz{\'a}lez-N{\'u}{\~n}ez}, J. and {Gonz{\'a}lez-Santamar{\'\i}a}, I. and {Gonz{\'a}lez-Vidal}, J.~J. and {Granvik}, M. and {Guillout}, P. and {Guiraud}, J. and {Guti{\'e}rrez-S{\'a}nchez}, R. and {Guy}, L.~P. and {Hatzidimitriou}, D. and {Hauser}, M. and {Haywood}, M. and {Helmer}, A. and {Helmi}, A. and {Sarmiento}, M.~H. and {Hidalgo}, S.~L. and {Hilger}, T. and {H{\l}adczuk}, N. and {Hobbs}, D. and {Holland}, G. and {Huckle}, H.~E. and {Jardine}, K. and {Jasniewicz}, G. and {Jean-Antoine Piccolo}, A. and {Jim{\'e}nez-Arranz}, {\'O}. and {Jorissen}, A. and {Juaristi Campillo}, J. and {Julbe}, F. and {Karbevska}, L. and {Kervella}, P. and {Khanna}, S. and {Kontizas}, M. and {Kordopatis}, G. and {Korn}, A.~J. and {K{\'o}sp{\'a}l}, {\'A}. and {Kostrzewa-Rutkowska}, Z. and {Kruszy{\'n}ska}, K. and {Kun}, M. and {Laizeau}, P. and {Lambert}, S. and {Lanza}, A.~F. and {Lasne}, Y. and {Le Campion}, J. -F. and {Lebreton}, Y. and {Lebzelter}, T. and {Leccia}, S. and {Leclerc}, N. and {Lecoeur-Taibi}, I. and {Liao}, S. and {Licata}, E.~L. and {Lindstr{\o}m}, H.~E.~P. and {Lister}, T.~A. and {Livanou}, E. and {Lobel}, A. and {Lorca}, A. and {Loup}, C. and {Madrero Pardo}, P. and {Magdaleno Romeo}, A. and {Managau}, S. and {Mann}, R.~G. and {Manteiga}, M. and {Marchant}, J.~M. and {Marconi}, M. and {Marcos}, J. and {Marcos Santos}, M.~M.~S. and {Mar{\'\i}n Pina}, D. and {Marinoni}, S. and {Marocco}, F. and {Marshall}, D.~J. and {Martin Polo}, L. and {Mart{\'\i}n-Fleitas}, J.~M. and {Marton}, G. and {Mary}, N. and {Masip}, A. and {Massari}, D. and {Mastrobuono-Battisti}, A. and {Mazeh}, T. and {McMillan}, P.~J. and {Messina}, S. and {Michalik}, D. and {Millar}, N.~R. and {Mints}, A. and {Molina}, D. and {Molinaro}, R. and {Moln{\'a}r}, L. and {Monari}, G. and {Mongui{\'o}}, M. and {Montegriffo}, P. and {Montero}, A. and {Mor}, R. and {Mora}, A. and {Morbidelli}, R. and {Morel}, T. and {Morris}, D. and {Muraveva}, T. and {Murphy}, C.~P. and {Musella}, I. and {Nagy}, Z. and {Noval}, L. and {Oca{\~n}a}, F. and {Ogden}, A. and {Ordenovic}, C. and {Osinde}, J.~O. and {Pagani}, C. and {Pagano}, I. and {Palaversa}, L. and {Palicio}, P.~A. and {Pallas-Quintela}, L. and {Panahi}, A. and {Payne-Wardenaar}, S. and {Pe{\~n}alosa Esteller}, X. and {Penttil{\"a}}, A. and {Pichon}, B. and {Piersimoni}, A.~M. and {Pineau}, F. -X. and {Plachy}, E. and {Plum}, G. and {Poggio}, E. and {Pr{\v{s}}a}, A. and {Pulone}, L. and {Racero}, E. and {Ragaini}, S. and {Rainer}, M. and {Raiteri}, C.~M. and {Rambaux}, N. and {Ramos}, P. and {Ramos-Lerate}, M. and {Re Fiorentin}, P. and {Regibo}, S. and {Richards}, P.~J. and {Rios Diaz}, C. and {Ripepi}, V. and {Riva}, A. and {Rix}, H. -W. and {Rixon}, G. and {Robichon}, N. and {Robin}, A.~C. and {Robin}, C. and {Roelens}, M. and {Rogues}, H.~R.~O. and {Rohrbasser}, L. and {Romero-G{\'o}mez}, M. and {Rowell}, N. and {Royer}, F. and {Ruz Mieres}, D. and {Rybicki}, K.~A. and {Sadowski}, G. and {S{\'a}ez N{\'u}{\~n}ez}, A. and {Sagrist{\`a} Sell{\'e}s}, A. and {Sahlmann}, J. and {Salguero}, E. and {Samaras}, N. and {Sanchez Gimenez}, V. and {Sanna}, N. and {Santove{\~n}a}, R. and {Sarasso}, M. and {Schultheis}, M. and {Sciacca}, E. and {Segol}, M. and {Segovia}, J.~C. and {S{\'e}gransan}, D. and {Semeux}, D. and {Shahaf}, S. and {Siddiqui}, H.~I. and {Siebert}, A. and {Siltala}, L. and {Silvelo}, A. and {Slezak}, E. and {Slezak}, I. and {Smart}, R.~L. and {Snaith}, O.~N. and {Solano}, E. and {Solitro}, F. and {Souami}, D. and {Souchay}, J. and {Spagna}, A. and {Spina}, L. and {Spoto}, F. and {Steele}, I.~A. and {Steidelm{\"u}ller}, H. and {Stephenson}, C.~A. and {S{\"u}veges}, M. and {Surdej}, J. and {Szabados}, L. and {Szegedi-Elek}, E. and {Taris}, F. and {Taylor}, M.~B. and {Teixeira}, R. and {Tolomei}, L. and {Tonello}, N. and {Torra}, F. and {Torra}, J. and {Torralba Elipe}, G. and {Trabucchi}, M. and {Tsounis}, A.~T. and {Turon}, C. and {Ulla}, A. and {Unger}, N. and {Vaillant}, M.~V. and {van Dillen}, E. and {van Reeven}, W. and {Vanel}, O. and {Vecchiato}, A. and {Viala}, Y. and {Vicente}, D. and {Voutsinas}, S. and {Weiler}, M. and {Wevers}, T. and {Wyrzykowski}, {\L}. and {Yoldas}, A. and {Yvard}, P. and {Zhao}, H. and {Zorec}, J. and {Zucker}, S. and {Zwitter}, T.},
        title = "{Gaia Data Release 3. Summary of the content and survey properties}",
      journal = {\aap},
         year = 2023,
        month = jun,
       volume = {674},
          eid = {A1},
        pages = {A1},
          doi = {10.1051/0004-6361/202243940},
archivePrefix = {arXiv},
       eprint = {2208.00211},
 primaryClass = {astro-ph.GA},
       adsurl = {https://ui.adsabs.harvard.edu/abs/2023A&A...674A...1G}
}

@ARTICLE{2016arXiv161205560C,
       author = {{Chambers}, K.~C. and {Magnier}, E.~A. and {Metcalfe}, N. and {Flewelling}, H.~A. and {Huber}, M.~E. and {Waters}, C.~Z. and {Denneau}, L. and {Draper}, P.~W. and {Farrow}, D. and {Finkbeiner}, D.~P. and {Holmberg}, C. and {Koppenhoefer}, J. and {Price}, P.~A. and {Rest}, A. and {Saglia}, R.~P. and {Schlafly}, E.~F. and {Smartt}, S.~J. and {Sweeney}, W. and {Wainscoat}, R.~J. and {Burgett}, W.~S. and {Chastel}, S. and {Grav}, T. and {Heasley}, J.~N. and {Hodapp}, K.~W. and {Jedicke}, R. and {Kaiser}, N. and {Kudritzki}, R. -P. and {Luppino}, G.~A. and {Lupton}, R.~H. and {Monet}, D.~G. and {Morgan}, J.~S. and {Onaka}, P.~M. and {Shiao}, B. and {Stubbs}, C.~W. and {Tonry}, J.~L. and {White}, R. and {Ba{\~n}ados}, E. and {Bell}, E.~F. and {Bender}, R. and {Bernard}, E.~J. and {Boegner}, M. and {Boffi}, F. and {Botticella}, M.~T. and {Calamida}, A. and {Casertano}, S. and {Chen}, W. -P. and {Chen}, X. and {Cole}, S. and {Deacon}, N. and {Frenk}, C. and {Fitzsimmons}, A. and {Gezari}, S. and {Gibbs}, V. and {Goessl}, C. and {Goggia}, T. and {Gourgue}, R. and {Goldman}, B. and {Grant}, P. and {Grebel}, E.~K. and {Hambly}, N.~C. and {Hasinger}, G. and {Heavens}, A.~F. and {Heckman}, T.~M. and {Henderson}, R. and {Henning}, T. and {Holman}, M. and {Hopp}, U. and {Ip}, W. -H. and {Isani}, S. and {Jackson}, M. and {Keyes}, C.~D. and {Koekemoer}, A.~M. and {Kotak}, R. and {Le}, D. and {Liska}, D. and {Long}, K.~S. and {Lucey}, J.~R. and {Liu}, M. and {Martin}, N.~F. and {Masci}, G. and {McLean}, B. and {Mindel}, E. and {Misra}, P. and {Morganson}, E. and {Murphy}, D.~N.~A. and {Obaika}, A. and {Narayan}, G. and {Nieto-Santisteban}, M.~A. and {Norberg}, P. and {Peacock}, J.~A. and {Pier}, E.~A. and {Postman}, M. and {Primak}, N. and {Rae}, C. and {Rai}, A. and {Riess}, A. and {Riffeser}, A. and {Rix}, H.~W. and {R{\"o}ser}, S. and {Russel}, R. and {Rutz}, L. and {Schilbach}, E. and {Schultz}, A.~S.~B. and {Scolnic}, D. and {Strolger}, L. and {Szalay}, A. and {Seitz}, S. and {Small}, E. and {Smith}, K.~W. and {Soderblom}, D.~R. and {Taylor}, P. and {Thomson}, R. and {Taylor}, A.~N. and {Thakar}, A.~R. and {Thiel}, J. and {Thilker}, D. and {Unger}, D. and {Urata}, Y. and {Valenti}, J. and {Wagner}, J. and {Walder}, T. and {Walter}, F. and {Watters}, S.~P. and {Werner}, S. and {Wood-Vasey}, W.~M. and {Wyse}, R.},
        title = "{The Pan-STARRS1 Surveys}",
      journal = {arXiv e-prints},
         year = 2016,
        month = dec,
          eid = {arXiv:1612.05560},
        pages = {arXiv:1612.05560},
          doi = {10.48550/arXiv.1612.05560},
archivePrefix = {arXiv},
       eprint = {1612.05560},
 primaryClass = {astro-ph.IM},
       adsurl = {https://ui.adsabs.harvard.edu/abs/2016arXiv161205560C}
}

@ARTICLE{2masscatalog,
       author = {{Cutri}, R.~M. and {Skrutskie}, M.~F. and {van Dyk}, S. and {Beichman}, C.~A. and {Carpenter}, J.~M. and {Chester}, T. and {Cambresy}, L. and {Evans}, T. and {Fowler}, J. and {Gizis}, J. and {Howard}, E. and {Huchra}, J. and {Jarrett}, T. and {Kopan}, E.~L. and {Kirkpatrick}, J.~D. and {Light}, R.~M. and {Marsh}, K.~A. and {McCallon}, H. and {Schneider}, S. and {Stiening}, R. and {Sykes}, M. and {Weinberg}, M. and {Wheaton}, W.~A. and {Wheelock}, S. and {Zacarias}, N.},
        title = "{VizieR Online Data Catalog: 2MASS All-Sky Catalog of Point Sources (Cutri+ 2003)}",
      journal = {VizieR Online Data Catalog},
         year = 2003,
        month = jun,
          eid = {II/246},
        pages = {II/246},
       adsurl = {https://ui.adsabs.harvard.edu/abs/2003yCat.2246....0C}
}

@ARTICLE{galexcatalog,
       author = {{Bianchi}, Luciana and {Shiao}, Bernie and {Thilker}, David},
        title = "{Revised Catalog of GALEX Ultraviolet Sources. I. The All-Sky Survey: GUVcat\_AIS}",
      journal = {\apjs},
         year = 2017,
        month = jun,
       volume = {230},
       number = {2},
          eid = {24},
        pages = {24},
          doi = {10.3847/1538-4365/aa7053},
archivePrefix = {arXiv},
       eprint = {1704.05903},
 primaryClass = {astro-ph.GA},
       adsurl = {https://ui.adsabs.harvard.edu/abs/2017ApJS..230...24B}
}

@INPROCEEDINGS{apasscatalog,
       author = {{Henden}, Arne A. and {Levine}, Stephen and {Terrell}, Dirk and {Welch}, Douglas L.},
        title = "{APASS - The Latest Data Release}",
    booktitle = {American Astronomical Society Meeting Abstracts \#225},
         year = 2015,
       series = {American Astronomical Society Meeting Abstracts},
       volume = {225},
        month = jan,
          eid = {336.16},
        pages = {336.16},
       adsurl = {https://ui.adsabs.harvard.edu/abs/2015AAS...22533616H}
}

@ARTICLE{sdsscatalog,
       author = {{Alam}, Shadab and {Albareti}, Franco D. and {Allende Prieto}, Carlos and {Anders}, F. and {Anderson}, Scott F. and {Anderton}, Timothy and {Andrews}, Brett H. and {Armengaud}, Eric and {Aubourg}, {\'E}ric and {Bailey}, Stephen and {Basu}, Sarbani and {Bautista}, Julian E. and {Beaton}, Rachael L. and {Beers}, Timothy C. and {Bender}, Chad F. and {Berlind}, Andreas A. and {Beutler}, Florian and {Bhardwaj}, Vaishali and {Bird}, Jonathan C. and {Bizyaev}, Dmitry and {Blake}, Cullen H. and {Blanton}, Michael R. and {Blomqvist}, Michael and {Bochanski}, John J. and {Bolton}, Adam S. and {Bovy}, Jo and {Shelden Bradley}, A. and {Brandt}, W.~N. and {Brauer}, D.~E. and {Brinkmann}, J. and {Brown}, Peter J. and {Brownstein}, Joel R. and {Burden}, Angela and {Burtin}, Etienne and {Busca}, Nicol{\'a}s G. and {Cai}, Zheng and {Capozzi}, Diego and {Carnero Rosell}, Aurelio and {Carr}, Michael A. and {Carrera}, Ricardo and {Chambers}, K.~C. and {Chaplin}, William James and {Chen}, Yen-Chi and {Chiappini}, Cristina and {Chojnowski}, S. Drew and {Chuang}, Chia-Hsun and {Clerc}, Nicolas and {Comparat}, Johan and {Covey}, Kevin and {Croft}, Rupert A.~C. and {Cuesta}, Antonio J. and {Cunha}, Katia and {da Costa}, Luiz N. and {Da Rio}, Nicola and {Davenport}, James R.~A. and {Dawson}, Kyle S. and {De Lee}, Nathan and {Delubac}, Timoth{\'e}e and {Deshpande}, Rohit and {Dhital}, Saurav and {Dutra-Ferreira}, Let{\'\i}cia and {Dwelly}, Tom and {Ealet}, Anne and {Ebelke}, Garrett L. and {Edmondson}, Edward M. and {Eisenstein}, Daniel J. and {Ellsworth}, Tristan and {Elsworth}, Yvonne and {Epstein}, Courtney R. and {Eracleous}, Michael and {Escoffier}, Stephanie and {Esposito}, Massimiliano and {Evans}, Michael L. and {Fan}, Xiaohui and {Fern{\'a}ndez-Alvar}, Emma and {Feuillet}, Diane and {Filiz Ak}, Nurten and {Finley}, Hayley and {Finoguenov}, Alexis and {Flaherty}, Kevin and {Fleming}, Scott W. and {Font-Ribera}, Andreu and {Foster}, Jonathan and {Frinchaboy}, Peter M. and {Galbraith-Frew}, J.~G. and {Garc{\'\i}a}, Rafael A. and {Garc{\'\i}a-Hern{\'a}ndez}, D.~A. and {Garc{\'\i}a P{\'e}rez}, Ana E. and {Gaulme}, Patrick and {Ge}, Jian and {G{\'e}nova-Santos}, R. and {Georgakakis}, A. and {Ghezzi}, Luan and {Gillespie}, Bruce A. and {Girardi}, L{\'e}o and {Goddard}, Daniel and {Gontcho}, Satya Gontcho A. and {Gonz{\'a}lez Hern{\'a}ndez}, Jonay I. and {Grebel}, Eva K. and {Green}, Paul J. and {Grieb}, Jan Niklas and {Grieves}, Nolan and {Gunn}, James E. and {Guo}, Hong and {Harding}, Paul and {Hasselquist}, Sten and {Hawley}, Suzanne L. and {Hayden}, Michael and {Hearty}, Fred R. and {Hekker}, Saskia and {Ho}, Shirley and {Hogg}, David W. and {Holley-Bockelmann}, Kelly and {Holtzman}, Jon A. and {Honscheid}, Klaus and {Huber}, Daniel and {Huehnerhoff}, Joseph and {Ivans}, Inese I. and {Jiang}, Linhua and {Johnson}, Jennifer A. and {Kinemuchi}, Karen and {Kirkby}, David and {Kitaura}, Francisco and {Klaene}, Mark A. and {Knapp}, Gillian R. and {Kneib}, Jean-Paul and {Koenig}, Xavier P. and {Lam}, Charles R. and {Lan}, Ting-Wen and {Lang}, Dustin and {Laurent}, Pierre and {Le Goff}, Jean-Marc and {Leauthaud}, Alexie and {Lee}, Khee-Gan and {Lee}, Young Sun and {Licquia}, Timothy C. and {Liu}, Jian and {Long}, Daniel C. and {L{\'o}pez-Corredoira}, Mart{\'\i}n and {Lorenzo-Oliveira}, Diego and {Lucatello}, Sara and {Lundgren}, Britt and {Lupton}, Robert H. and {Mack}, Claude E., III and {Mahadevan}, Suvrath and {Maia}, Marcio A.~G. and {Majewski}, Steven R. and {Malanushenko}, Elena and {Malanushenko}, Viktor and {Manchado}, A. and {Manera}, Marc and {Mao}, Qingqing and {Maraston}, Claudia and {Marchwinski}, Robert C. and {Margala}, Daniel and {Martell}, Sarah L. and {Martig}, Marie and {Masters}, Karen L. and {Mathur}, Savita and {McBride}, Cameron K. and {McGehee}, Peregrine M. and {McGreer}, Ian D. and {McMahon}, Richard G. and {M{\'e}nard}, Brice and {Menzel}, Marie-Luise and {Merloni}, Andrea and {M{\'e}sz{\'a}ros}, Szabolcs and {Miller}, Adam A. and {Miralda-Escud{\'e}}, Jordi and {Miyatake}, Hironao and {Montero-Dorta}, Antonio D. and {More}, Surhud and {Morganson}, Eric and {Morice-Atkinson}, Xan and {Morrison}, Heather L. and {Mosser}, Ben{\^o}it and {Muna}, Demitri and {Myers}, Adam D. and {Nandra}, Kirpal and {Newman}, Jeffrey A. and {Neyrinck}, Mark and {Nguyen}, Duy Cuong and {Nichol}, Robert C. and {Nidever}, David L. and {Noterdaeme}, Pasquier and {Nuza}, Sebasti{\'a}n E. and {O'Connell}, Julia E. and {O'Connell}, Robert W. and {O'Connell}, Ross and {Ogando}, Ricardo L.~C. and {Olmstead}, Matthew D. and {Oravetz}, Audrey E. and {Oravetz}, Daniel J. and {Osumi}, Keisuke and {Owen}, Russell and {Padgett}, Deborah L. and {Padmanabhan}, Nikhil and {Paegert}, Martin and {Palanque-Delabrouille}, Nathalie and {Pan}, Kaike and {Parejko}, John K. and {P{\^a}ris}, Isabelle and {Park}, Changbom and {Pattarakijwanich}, Petchara and {Pellejero-Ibanez}, M. and {Pepper}, Joshua and {Percival}, Will J. and {P{\'e}rez-Fournon}, Ismael and {P{\'e}rez-R{\`a}fols}, Ignasi and {Petitjean}, Patrick and {Pieri}, Matthew M. and {Pinsonneault}, Marc H. and {Porto de Mello}, Gustavo F. and {Prada}, Francisco and {Prakash}, Abhishek and {Price-Whelan}, Adrian M. and {Protopapas}, Pavlos and {Raddick}, M. Jordan and {Rahman}, Mubdi and {Reid}, Beth A. and {Rich}, James and {Rix}, Hans-Walter and {Robin}, Annie C. and {Rockosi}, Constance M. and {Rodrigues}, Tha{\'\i}se S. and {Rodr{\'\i}guez-Torres}, Sergio and {Roe}, Natalie A. and {Ross}, Ashley J. and {Ross}, Nicholas P. and {Rossi}, Graziano and {Ruan}, John J. and {Rubi{\~n}o-Mart{\'\i}n}, J.~A. and {Rykoff}, Eli S. and {Salazar-Albornoz}, Salvador and {Salvato}, Mara and {Samushia}, Lado and {S{\'a}nchez}, Ariel G. and {Santiago}, Bas{\'\i}lio and {Sayres}, Conor and {Schiavon}, Ricardo P. and {Schlegel}, David J. and {Schmidt}, Sarah J. and {Schneider}, Donald P. and {Schultheis}, Mathias and {Schwope}, Axel D. and {Sc{\'o}ccola}, C.~G. and {Scott}, Caroline and {Sellgren}, Kris and {Seo}, Hee-Jong and {Serenelli}, Aldo and {Shane}, Neville and {Shen}, Yue and {Shetrone}, Matthew and {Shu}, Yiping and {Silva Aguirre}, V. and {Sivarani}, Thirupathi and {Skrutskie}, M.~F. and {Slosar}, An{\v{z}}e and {Smith}, Verne V. and {Sobreira}, Fl{\'a}via and {Souto}, Diogo and {Stassun}, Keivan G. and {Steinmetz}, Matthias and {Stello}, Dennis and {Strauss}, Michael A. and {Streblyanska}, Alina and {Suzuki}, Nao and {Swanson}, Molly E.~C. and {Tan}, Jonathan C. and {Tayar}, Jamie and {Terrien}, Ryan C. and {Thakar}, Aniruddha R. and {Thomas}, Daniel and {Thomas}, Neil and {Thompson}, Benjamin A. and {Tinker}, Jeremy L. and {Tojeiro}, Rita and {Troup}, Nicholas W. and {Vargas-Maga{\~n}a}, Mariana and {Vazquez}, Jose A. and {Verde}, Licia and {Viel}, Matteo and {Vogt}, Nicole P. and {Wake}, David A. and {Wang}, Ji and {Weaver}, Benjamin A. and {Weinberg}, David H. and {Weiner}, Benjamin J. and {White}, Martin and {Wilson}, John C. and {Wisniewski}, John P. and {Wood-Vasey}, W.~M. and {Ye`che}, Christophe and {York}, Donald G. and {Zakamska}, Nadia L. and {Zamora}, O. and {Zasowski}, Gail and {Zehavi}, Idit and {Zhao}, Gong-Bo and {Zheng}, Zheng and {Zhou}, Xu and {Zhou}, Zhimin and {Zou}, Hu and {Zhu}, Guangtun},
        title = "{The Eleventh and Twelfth Data Releases of the Sloan Digital Sky Survey: Final Data from SDSS-III}",
      journal = {\apjs},
         year = 2015,
        month = jul,
       volume = {219},
       number = {1},
          eid = {12},
        pages = {12},
          doi = {10.1088/0067-0049/219/1/12},
archivePrefix = {arXiv},
       eprint = {1501.00963},
 primaryClass = {astro-ph.IM},
       adsurl = {https://ui.adsabs.harvard.edu/abs/2015ApJS..219...12A}
}

@ARTICLE{allwisecatalog,
       author = {{Cutri}, R.~M. and {Wright}, E.~L. and {Conrow}, T. and {Fowler}, J.~W. and {Eisenhardt}, P.~R.~M. and {Grillmair}, C. and {Kirkpatrick}, J.~D. and {Masci}, F. and {McCallon}, H.~L. and {Wheelock}, S.~L. and {Fajardo-Acosta}, S. and {Yan}, L. and {Benford}, D. and {Harbut}, M. and {Jarrett}, T. and {Lake}, S. and {Leisawitz}, D. and {Ressler}, M.~E. and {Stanford}, S.~A. and {Tsai}, C. -W. and {Liu}, F. and {Helou}, G. and {Mainzer}, A. and {Gettngs}, D. and {Gonzalez}, A. and {Hoffman}, D. and {Marsh}, K.~A. and {Padgett}, D. and {Skrutskie}, M.~F. and {Beck}, R. and {Papin}, M. and {Wittman}, M.},
        title = "{VizieR Online Data Catalog: AllWISE Data Release (Cutri+ 2013)}",
      journal = {VizieR Online Data Catalog},
         year = 2021,
        month = feb,
          eid = {II/328},
        pages = {II/328},
       adsurl = {https://ui.adsabs.harvard.edu/abs/2014yCat.2328....0C}
}

@ARTICLE{2021MNRAS.507.2373P,
       author = {{Pandey}, Sindhu and {Subramaniam}, Annapurni and {Jadhav}, Vikrant V.},
        title = "{UOCS - VI. UVIT/AstroSat detection of low-mass white dwarf companions to four more blue stragglers in M67}",
      journal = {\mnras},
         year = 2021,
        month = oct,
       volume = {507},
       number = {2},
        pages = {2373-2382},
          doi = {10.1093/mnras/stab2308},
archivePrefix = {arXiv},
       eprint = {2108.03371},
 primaryClass = {astro-ph.SR},
       adsurl = {https://ui.adsabs.harvard.edu/abs/2021MNRAS.507.2373P}
}

@ARTICLE{2006A&A...451..993C,
       author = {{Canto Martins}, B.~L. and {L{\`e}bre}, A. and {de Laverny}, P. and {Melo}, C.~H.~F. and {Do Nascimento}, Jr., J.~D. and {Richard}, O. and {de Medeiros}, J.~R.},
        title = "{S1242: a lithium-rich subgiant star in the open cluster M 67}",
      journal = {\aap},
         year = 2006,
        month = jun,
       volume = {451},
       number = {3},
        pages = {993-997},
          doi = {10.1051/0004-6361:20053334},
       adsurl = {https://ui.adsabs.harvard.edu/abs/2006A&A...451..993C}
}

@ARTICLE{2011A&A...527A..94C,
       author = {{Canto Martins}, B.~L. and {L{\`e}bre}, A. and {Palacios}, A. and {de Laverny}, P. and {Richard}, O. and {Melo}, C.~H.~F. and {Do Nascimento}, Jr., J.~D. and {de Medeiros}, J.~R.},
        title = "{Lithium abundances and extra mixing processes in evolved stars of M 67}",
      journal = {\aap},
         year = 2011,
        month = mar,
       volume = {527},
          eid = {A94},
        pages = {A94},
          doi = {10.1051/0004-6361/201015015},
archivePrefix = {arXiv},
       eprint = {1101.3757},
 primaryClass = {astro-ph.SR},
       adsurl = {https://ui.adsabs.harvard.edu/abs/2011A&A...527A..94C}
}

@ARTICLE{2012A&A...541A.150P,
       author = {{Pace}, G. and {Castro}, M. and {Mel{\'e}ndez}, J. and {Th{\'e}ado}, S. and {do Nascimento}, Jr., J. -D.},
        title = "{Lithium in M 67: From the main sequence to the red giant branch}",
      journal = {\aap},
         year = 2012,
        month = may,
       volume = {541},
          eid = {A150},
        pages = {A150},
          doi = {10.1051/0004-6361/201117704},
archivePrefix = {arXiv},
       eprint = {1203.4440},
 primaryClass = {astro-ph.SR},
       adsurl = {https://ui.adsabs.harvard.edu/abs/2012A&A...541A.150P}
}

@ARTICLE{2004A&A...418..509V,
       author = {{van den Berg}, M. and {Tagliaferri}, G. and {Belloni}, T. and {Verbunt}, F.},
        title = "{A Chandra observation of the old open cluster M 67}",
      journal = {\aap},
         year = 2004,
        month = may,
       volume = {418},
        pages = {509-523},
          doi = {10.1051/0004-6361:20031642},
archivePrefix = {arXiv},
       eprint = {astro-ph/0311138},
 primaryClass = {astro-ph},
       adsurl = {https://ui.adsabs.harvard.edu/abs/2004A&A...418..509V}
}

@ARTICLE{1999A&A...347..866V,
       author = {{van den Berg}, M. and {Verbunt}, F. and {Mathieu}, R.~D.},
        title = "{Optical spectroscopy of X-ray sources in the old open cluster M 67}",
      journal = {\aap},
         year = 1999,
        month = jul,
       volume = {347},
        pages = {866-875},
          doi = {10.48550/arXiv.astro-ph/9904217},
archivePrefix = {arXiv},
       eprint = {astro-ph/9904217},
 primaryClass = {astro-ph},
       adsurl = {https://ui.adsabs.harvard.edu/abs/1999A&A...347..866V}
}

@ARTICLE{1998A&A...339..431B,
       author = {{Belloni}, T. and {Verbunt}, F. and {Mathieu}, R.~D.},
        title = "{X-rays from old open clusters: M 67 and NGC 188}",
      journal = {\aap},
         year = 1998,
        month = nov,
       volume = {339},
        pages = {431-439},
          doi = {10.48550/arXiv.astro-ph/9808329},
archivePrefix = {arXiv},
       eprint = {astro-ph/9808329},
 primaryClass = {astro-ph},
       adsurl = {https://ui.adsabs.harvard.edu/abs/1998A&A...339..431B}
}

@ARTICLE{1990AJ....100.1859M,
       author = {{Mathieu}, Robert D. and {Latham}, David W. and {Griffin}, R.~F.},
        title = "{Orbits of 22 Spectroscopic Binaries in the Open Cluster M67}",
      journal = {\aj},
         year = 1990,
        month = dec,
       volume = {100},
        pages = {1859},
          doi = {10.1086/115643},
       adsurl = {https://ui.adsabs.harvard.edu/abs/1990AJ....100.1859M}
}

@ARTICLE{2015AJ....150...97G,
       author = {{Geller}, Aaron M. and {Latham}, David W. and {Mathieu}, Robert D.},
        title = "{Stellar Radial Velocities in the Old Open Cluster M67 (NGC 2682). I. Memberships, Binaries, and Kinematics}",
      journal = {\aj},
         year = 2015,
        month = sep,
       volume = {150},
       number = {3},
          eid = {97},
        pages = {97},
          doi = {10.1088/0004-6256/150/3/97},
archivePrefix = {arXiv},
       eprint = {1507.01949},
 primaryClass = {astro-ph.SR},
       adsurl = {https://ui.adsabs.harvard.edu/abs/2015AJ....150...97G}
}

@ARTICLE{2021AJ....161..190G,
       author = {{Geller}, Aaron M. and {Mathieu}, Robert D. and {Latham}, David W. and {Pollack}, Maxwell and {Torres}, Guillermo and {Leiner}, Emily M.},
        title = "{Stellar Radial Velocities in the Old Open Cluster M67 (NGC 2682). II. The Spectroscopic Binary Population}",
      journal = {\aj},
         year = 2021,
        month = apr,
       volume = {161},
       number = {4},
          eid = {190},
        pages = {190},
          doi = {10.3847/1538-3881/abdd23},
archivePrefix = {arXiv},
       eprint = {2101.07883},
 primaryClass = {astro-ph.SR},
       adsurl = {https://ui.adsabs.harvard.edu/abs/2021AJ....161..190G}
}

@ARTICLE{2018MNRAS.481..226S,
       author = {{Sindhu}, N. and {Subramaniam}, Annapurni and {Radha}, C. Anu},
        title = "{Ultraviolet stellar population of the old open cluster M67 (NGC 2682)}",
      journal = {\mnras},
         year = 2018,
        month = nov,
       volume = {481},
       number = {1},
        pages = {226-243},
          doi = {10.1093/mnras/sty2283},
archivePrefix = {arXiv},
       eprint = {1808.06814},
 primaryClass = {astro-ph.SR},
       adsurl = {https://ui.adsabs.harvard.edu/abs/2018MNRAS.481..226S}
}

@ARTICLE{2023A&A...676A.129H,
       author = {{Hourihane}, A. and {Fran{\c{c}}ois}, P. and {Worley}, C.~C. and {Magrini}, L. and {Gonneau}, A. and {Casey}, A.~R. and {Gilmore}, G. and {Randich}, S. and {Sacco}, G.~G. and {Recio-Blanco}, A. and {Korn}, A.~J. and {Allende Prieto}, C. and {Smiljanic}, R. and {Blomme}, R. and {Bragaglia}, A. and {Walton}, N.~A. and {Van Eck}, S. and {Bensby}, T. and {Lanzafame}, A. and {Frasca}, A. and {Franciosini}, E. and {Damiani}, F. and {Lind}, K. and {Bergemann}, M. and {Bonifacio}, P. and {Hill}, V. and {Lobel}, A. and {Montes}, D. and {Feuillet}, D.~K. and {Tautvai{\v{s}}ien{\.{e}}}, G. and {Guiglion}, G. and {Tabernero}, H.~M. and {Gonz{\'a}lez Hern{\'a}ndez}, J.~I. and {Gebran}, M. and {Van der Swaelmen}, M. and {Mikolaitis}, {\v{S}}. and {Daflon}, S. and {Merle}, T. and {Morel}, T. and {Lewis}, J.~R. and {Gonz{\'a}lez Solares}, E.~A. and {Murphy}, D.~N.~A. and {Jeffries}, R.~D. and {Jackson}, R.~J. and {Feltzing}, S. and {Prusti}, T. and {Carraro}, G. and {Biazzo}, K. and {Prisinzano}, L. and {Jofr{\'e}}, P. and {Zaggia}, S. and {Drazdauskas}, A. and {Stonkut{\'e}}, E. and {Marfil}, E. and {Jim{\'e}nez-Esteban}, F. and {Mahy}, L. and {Guti{\'e}rrez Albarr{\'a}n}, M.~L. and {Berlanas}, S.~R. and {Santos}, W. and {Morbidelli}, L. and {Spina}, L. and {Minkevi{\v{c}}i{\={u}}t{\.{e}}}, R.},
        title = "{The Gaia-ESO Survey: Homogenisation of stellar parameters and elemental abundances}",
      journal = {\aap},
         year = 2023,
        month = aug,
       volume = {676},
          eid = {A129},
        pages = {A129},
          doi = {10.1051/0004-6361/202345910},
archivePrefix = {arXiv},
       eprint = {2304.07720},
 primaryClass = {astro-ph.SR},
       adsurl = {https://ui.adsabs.harvard.edu/abs/2023A&A...676A.129H}
}

@ARTICLE{2023A&A...674A.157T,
       author = {{Tsantaki}, M. and {Delgado-Mena}, E. and {Bossini}, D. and {Sousa}, S.~G. and {Pancino}, E. and {Martins}, J.~H.~C.},
        title = "{Search for lithium-rich giants in 32 open clusters with high-resolution spectroscopy★}",
      journal = {\aap},
         year = 2023,
        month = jun,
       volume = {674},
          eid = {A157},
        pages = {A157},
          doi = {10.1051/0004-6361/202244374},
archivePrefix = {arXiv},
       eprint = {2303.16124},
 primaryClass = {astro-ph.SR},
       adsurl = {https://ui.adsabs.harvard.edu/abs/2023A&A...674A.157T}
}

@ARTICLE{1997A&A...318..841C,
       author = {{Castelli}, F. and {Gratton}, R.~G. and {Kurucz}, R.~L.},
        title = "{Notes on the convection in the ATLAS9 model atmospheres.}",
      journal = {\aap},
         year = "1997",
        month = "Feb",
       volume = {318},
        pages = {841-869},
       adsurl = {https://ui.adsabs.harvard.edu/abs/1997A&A...318..841C}
}

@ARTICLE{2023ApJ...945...11R,
       author = {{Rani}, Sharmila and {Pandey}, Gajendra and {Subramaniam}, Annapurni and {Rao}, N. Kameswara},
        title = "{UOCS-IX. AstroSat/UVIT Study of the Open Cluster NGC 2818: Blue Stragglers, Yellow Stragglers, Planetary Nebula, and their Membership}",
      journal = {\apj},
         year = 2023,
        month = mar,
       volume = {945},
       number = {1},
          eid = {11},
        pages = {11},
          doi = {10.3847/1538-4357/acb0c8},
archivePrefix = {arXiv},
       eprint = {2301.01943},
 primaryClass = {astro-ph.SR},
       adsurl = {https://ui.adsabs.harvard.edu/abs/2023ApJ...945...11R}
}

@ARTICLE{2009ApJ...696.1755T,
       author = {{Tremblay}, P. -E. and {Bergeron}, P.},
        title = "{Spectroscopic Analysis of DA White Dwarfs: Stark Broadening of Hydrogen Lines Including Nonideal Effects}",
      journal = {\apj},
         year = 2009,
        month = may,
       volume = {696},
       number = {2},
        pages = {1755-1770},
          doi = {10.1088/0004-637X/696/2/1755},
archivePrefix = {arXiv},
       eprint = {0902.4182},
 primaryClass = {astro-ph.SR},
       adsurl = {https://ui.adsabs.harvard.edu/abs/2009ApJ...696.1755T}
}

@ARTICLE{2010MmSAI..81..921K,
       author = {{Koester}, D.},
        title = "{White dwarf spectra and atmosphere models}",
      journal = {\memsai},
         year = 2010,
        month = jan,
       volume = {81},
        pages = {921-931},
       adsurl = {https://ui.adsabs.harvard.edu/abs/2010MmSAI..81..921K}
}

@ARTICLE{2013A&A...557A..19A,
       author = {{Althaus}, Leandro G. and {Miller Bertolami}, Marcelo M. and {C{\'o}rsico}, Alejandro H.},
        title = "{New evolutionary sequences for extremely low-mass white dwarfs. Homogeneous mass and age determinations and asteroseismic prospects}",
      journal = {\aap},
         year = 2013,
        month = sep,
       volume = {557},
          eid = {A19},
        pages = {A19},
          doi = {10.1051/0004-6361/201321868},
archivePrefix = {arXiv},
       eprint = {1307.1882},
 primaryClass = {astro-ph.SR},
       adsurl = {https://ui.adsabs.harvard.edu/abs/2013A&A...557A..19A}
}

@ARTICLE{2023AJ....165...70C,
       author = {{Crandall}, Sara and {Smith}, Graeme H.},
        title = "{Correlations in Chromospheric and Coronal Activity Indicators of Giant Stars}",
      journal = {\aj},
         year = 2023,
        month = feb,
       volume = {165},
       number = {2},
          eid = {70},
        pages = {70},
          doi = {10.3847/1538-3881/aca150},
       adsurl = {https://ui.adsabs.harvard.edu/abs/2023AJ....165...70C}
}

@ARTICLE{2018PASA...35...37S,
       author = {{Smith}, Graeme H.},
        title = "{A correlation between GALEX FUV magnitude and chromospheric activity among red giant stars}",
      journal = {\pasa},
         year = 2018,
        month = nov,
       volume = {35},
          eid = {e037},
        pages = {e037},
          doi = {10.1017/pasa.2018.45},
       adsurl = {https://ui.adsabs.harvard.edu/abs/2018PASA...35...37S}
}

@ARTICLE{2020AJ....160...12D,
       author = {{Dixon}, Don and {Tayar}, Jamie and {Stassun}, Keivan G.},
        title = "{Rotationally Driven Ultraviolet Emission of Red Giant Stars}",
      journal = {\aj},
         year = 2020,
        month = jul,
       volume = {160},
       number = {1},
          eid = {12},
        pages = {12},
          doi = {10.3847/1538-3881/ab9080},
archivePrefix = {arXiv},
       eprint = {2005.00577},
 primaryClass = {astro-ph.SR},
       adsurl = {https://ui.adsabs.harvard.edu/abs/2020AJ....160...12D}
}

@ARTICLE{2025AJ....169..309D,
       author = {{Dixon}, D. and {Stassun}, K.~G. and {Mathieu}, R.~D. and {Tayar}, J. and {Cao}, L.},
        title = "{Rotationally Driven Ultraviolet Emission of Red Giant Stars. II. Metallicity, Activity, Binarity, and Subsubgiants.}",
      journal = {\aj},
         year = 2025,
        month = jan,
       volume = {169},
          eid = {309},
        pages = {309},
          doi = {10.3847/1538-3881/adc92a},
archivePrefix = {arXiv},
       eprint = {2504.05561},
 primaryClass = {astro-ph.SR},
       adsurl = {https://ui.adsabs.harvard.edu/abs/2025AJ....169..309D}
}

@ARTICLE{2015MNRAS.452.3394M,
       author = {{Mooley}, K.~P. and {Singh}, K.~P.},
        title = "{Study of X-ray emission from the old open cluster, M67}",
      journal = {\mnras},
         year = 2015,
        month = oct,
       volume = {452},
       number = {4},
        pages = {3394-3407},
          doi = {10.1093/mnras/stv1472},
archivePrefix = {arXiv},
       eprint = {1507.00759},
 primaryClass = {astro-ph.HE},
       adsurl = {https://ui.adsabs.harvard.edu/abs/2015MNRAS.452.3394M}
}

@ARTICLE{1998A&A...336..902P,
       author = {{Pasquini}, L. and {Belloni}, T.},
        title = "{Optical identification of ROSAT sources in M 67: activity in an old cluster}",
      journal = {\aap},
         year = 1998,
        month = aug,
       volume = {336},
        pages = {902-910},
          doi = {10.48550/arXiv.astro-ph/9805364},
archivePrefix = {arXiv},
       eprint = {astro-ph/9805364},
 primaryClass = {astro-ph},
       adsurl = {https://ui.adsabs.harvard.edu/abs/1998A&A...336..902P}
}

@ARTICLE{2023MNRAS.518.4579P,
       author = {{Parsons}, S.~G. and {Hernandez}, M.~S. and {Toloza}, O. and {Zorotovic}, M. and {Schreiber}, M.~R. and {G{\"a}nsicke}, B.~T. and {Lagos}, F. and {Raddi}, R. and {Rebassa-Mansergas}, A. and {Ren}, J.~J. and {Koester}, D.},
        title = "{The white dwarf binary pathways survey - IX. Three long period white dwarf plus subgiant binaries}",
      journal = {\mnras},
         year = 2023,
        month = jan,
       volume = {518},
       number = {3},
        pages = {4579-4594},
          doi = {10.1093/mnras/stac3368},
archivePrefix = {arXiv},
       eprint = {2211.08440},
 primaryClass = {astro-ph.SR},
       adsurl = {https://ui.adsabs.harvard.edu/abs/2023MNRAS.518.4579P}
}

@ARTICLE{2016MNRAS.463.2125P,
       author = {{Parsons}, S.~G. and {Rebassa-Mansergas}, A. and {Schreiber}, M.~R. and {G{\"a}nsicke}, B.~T. and {Zorotovic}, M. and {Ren}, J.~J.},
        title = "{The white dwarf binary pathways survey - I. A sample of FGK stars with white dwarf companions}",
      journal = {\mnras},
         year = 2016,
        month = dec,
       volume = {463},
       number = {2},
        pages = {2125-2136},
          doi = {10.1093/mnras/stw2143},
archivePrefix = {arXiv},
       eprint = {1604.01613},
 primaryClass = {astro-ph.SR},
       adsurl = {https://ui.adsabs.harvard.edu/abs/2016MNRAS.463.2125P}
}

@ARTICLE{2020A&A...642A.228W,
       author = {{Werner}, Klaus and {Reindl}, Nicole and {L{\"o}bling}, Lisa and {Pelisoli}, Ingrid and {Schaffenroth}, Veronika and {Rebassa-Mansergas}, Alberto and {Irawati}, Puji and {Ren}, Juanjuan},
        title = "{An extremely hot white dwarf with a rapidly rotating K-type subgiant companion: UCAC2 46706450}",
      journal = {\aap},
         year = 2020,
        month = oct,
       volume = {642},
          eid = {A228},
        pages = {A228},
          doi = {10.1051/0004-6361/202038574},
archivePrefix = {arXiv},
       eprint = {2009.02968},
 primaryClass = {astro-ph.SR},
       adsurl = {https://ui.adsabs.harvard.edu/abs/2020A&A...642A.228W}
}

@ARTICLE{2024A&A...689A..89M,
       author = {{Mucciarelli}, A. and {Bonifacio}, P. and {Monaco}, L. and {Salaris}, M. and {Matteuzzi}, M.},
        title = "{The true nature of HE 0057-5959, the most metal-poor, Li-rich star}",
      journal = {\aap},
         year = 2024,
        month = sep,
       volume = {689},
          eid = {A89},
        pages = {A89},
          doi = {10.1051/0004-6361/202449290},
archivePrefix = {arXiv},
       eprint = {2406.17026},
 primaryClass = {astro-ph.SR},
       adsurl = {https://ui.adsabs.harvard.edu/abs/2024A&A...689A..89M}
}

@ARTICLE{2020A&A...639L...2S,
       author = {{Sanna}, N. and {Franciosini}, E. and {Pancino}, E. and {Mucciarelli}, A. and {Tsantaki}, M. and {Charbonnel}, C. and {Smiljanic}, R. and {Fu}, X. and {Bragaglia}, A. and {Lagarde}, N. and {Tautvai{\v{s}}iene}, G. and {Magrini}, L. and {Randich}, S. and {Bensby}, T. and {Korn}, A.~J. and {Bayo}, A. and {Bergemann}, M. and {Carraro}, G. and {Morbidelli}, L.},
        title = "{The Gaia-ESO Survey: an extremely Li-rich giant in globular cluster NGC 1261}",
      journal = {\aap},
         year = 2020,
        month = jul,
       volume = {639},
          eid = {L2},
        pages = {L2},
          doi = {10.1051/0004-6361/202038435},
archivePrefix = {arXiv},
       eprint = {2006.04393},
 primaryClass = {astro-ph.SR},
       adsurl = {https://ui.adsabs.harvard.edu/abs/2020A&A...639L...2S}
}

@ARTICLE{1982ApJ...255..577W,
       author = {{Wallerstein}, G. and {Sneden}, C.},
        title = "{A K giant with an unusually high abundance of lithium : HD 112127.}",
      journal = {\apj},
         year = 1982,
        month = apr,
       volume = {255},
        pages = {577-584},
          doi = {10.1086/159859},
       adsurl = {https://ui.adsabs.harvard.edu/abs/1982ApJ...255..577W}
}

@ARTICLE{2018ApJ...852L..31L,
       author = {{Li}, Haining and {Aoki}, Wako and {Matsuno}, Tadafumi and {Bharat Kumar}, Yerra and {Shi}, Jianrong and {Suda}, Takuma and {Zhao}, Gang},
        title = "{Enormous Li Enhancement Preceding Red Giant Phases in Low-mass Stars in the Milky Way Halo}",
      journal = {\apjl},
         year = 2018,
        month = jan,
       volume = {852},
       number = {2},
          eid = {L31},
        pages = {L31},
          doi = {10.3847/2041-8213/aaa438},
archivePrefix = {arXiv},
       eprint = {1801.00090},
 primaryClass = {astro-ph.SR},
       adsurl = {https://ui.adsabs.harvard.edu/abs/2018ApJ...852L..31L}
}

@ARTICLE{2016MNRAS.461.3336C,
       author = {{Casey}, A.~R. and {Ruchti}, G. and {Masseron}, T. and {Randich}, S. and {Gilmore}, G. and {Lind}, K. and {Kennedy}, G.~M. and {Koposov}, S.~E. and {Hourihane}, A. and {Franciosini}, E. and {Lewis}, J.~R. and {Magrini}, L. and {Morbidelli}, L. and {Sacco}, G.~G. and {Worley}, C.~C. and {Feltzing}, S. and {Jeffries}, R.~D. and {Vallenari}, A. and {Bensby}, T. and {Bragaglia}, A. and {Flaccomio}, E. and {Francois}, P. and {Korn}, A.~J. and {Lanzafame}, A. and {Pancino}, E. and {Recio-Blanco}, A. and {Smiljanic}, R. and {Carraro}, G. and {Costado}, M.~T. and {Damiani}, F. and {Donati}, P. and {Frasca}, A. and {Jofr{\'e}}, P. and {Lardo}, C. and {de Laverny}, P. and {Monaco}, L. and {Prisinzano}, L. and {Sbordone}, L. and {Sousa}, S.~G. and {Tautvai{\v{s}}ien{\.{e}}}, G. and {Zaggia}, S. and {Zwitter}, T. and {Delgado Mena}, E. and {Chorniy}, Y. and {Martell}, S.~L. and {Silva Aguirre}, V. and {Miglio}, A. and {Chiappini}, C. and {Montalban}, J. and {Morel}, T. and {Valentini}, M.},
        title = "{The Gaia-ESO Survey: revisiting the Li-rich giant problem}",
      journal = {\mnras},
         year = 2016,
        month = sep,
       volume = {461},
       number = {3},
        pages = {3336-3352},
          doi = {10.1093/mnras/stw1512},
archivePrefix = {arXiv},
       eprint = {1603.03038},
 primaryClass = {astro-ph.SR},
       adsurl = {https://ui.adsabs.harvard.edu/abs/2016MNRAS.461.3336C}
}

@ARTICLE{2020MNRAS.494.1348D,
       author = {{Deepak} and {Lambert}, David L. and {Reddy}, Bacham E.},
        title = "{Abundance analyses of Li-enriched and normal giants in the GALAH survey}",
      journal = {\mnras},
         year = 2020,
        month = may,
       volume = {494},
       number = {1},
        pages = {1348-1365},
          doi = {10.1093/mnras/staa729},
archivePrefix = {arXiv},
       eprint = {2003.05381},
 primaryClass = {astro-ph.SR},
       adsurl = {https://ui.adsabs.harvard.edu/abs/2020MNRAS.494.1348D}
}

@ARTICLE{2012ApJ...752L..16K,
       author = {{Kirby}, Evan N. and {Fu}, Xiaoting and {Guhathakurta}, Puragra and {Deng}, Licai},
        title = "{Discovery of Super-Li-rich Red Giants in Dwarf Spheroidal Galaxies}",
      journal = {\apjl},
         year = 2012,
        month = jun,
       volume = {752},
       number = {1},
          eid = {L16},
        pages = {L16},
          doi = {10.1088/2041-8205/752/1/L16},
archivePrefix = {arXiv},
       eprint = {1205.1057},
 primaryClass = {astro-ph.SR},
       adsurl = {https://ui.adsabs.harvard.edu/abs/2012ApJ...752L..16K}
}

@ARTICLE{2016ApJ...819..135K,
       author = {{Kirby}, Evan N. and {Guhathakurta}, Puragra and {Zhang}, Andrew J. and {Hong}, Jerry and {Guo}, Michelle and {Guo}, Rachel and {Cohen}, Judith G. and {Cunha}, Katia},
        title = "{Lithium-rich Giants in Globular Clusters}",
      journal = {\apj},
         year = 2016,
        month = mar,
       volume = {819},
       number = {2},
          eid = {135},
        pages = {135},
          doi = {10.3847/0004-637X/819/2/135},
archivePrefix = {arXiv},
       eprint = {1601.01315},
 primaryClass = {astro-ph.SR},
       adsurl = {https://ui.adsabs.harvard.edu/abs/2016ApJ...819..135K}
}

@ARTICLE{2019A&A...623A..55M,
       author = {{Mucciarelli}, A. and {Monaco}, L. and {Bonifacio}, P. and {Salaris}, M. and {Fu}, X. and {Villanova}, S.},
        title = "{The discovery of a Li/Na-rich giant star in Omega Centauri: formed from the pure ejecta of super-AGB stars?}",
      journal = {\aap},
         year = 2019,
        month = mar,
       volume = {623},
          eid = {A55},
        pages = {A55},
          doi = {10.1051/0004-6361/201834497},
archivePrefix = {arXiv},
       eprint = {1901.00048},
 primaryClass = {astro-ph.SR},
       adsurl = {https://ui.adsabs.harvard.edu/abs/2019A&A...623A..55M}
}

@ARTICLE{2014A&A...564L...6M,
       author = {{Monaco}, L. and {Boffin}, H.~M.~J. and {Bonifacio}, P. and {Villanova}, S. and {Carraro}, G. and {Caffau}, E. and {Steffen}, M. and {Ahumada}, J.~A. and {Beletsky}, Y. and {Beccari}, G.},
        title = "{A super lithium-rich red-clump star in the open cluster Trumpler 5}",
      journal = {\aap},
         year = 2014,
        month = apr,
       volume = {564},
          eid = {L6},
        pages = {L6},
          doi = {10.1051/0004-6361/201323348},
archivePrefix = {arXiv},
       eprint = {1403.6461},
 primaryClass = {astro-ph.SR},
       adsurl = {https://ui.adsabs.harvard.edu/abs/2014A&A...564L...6M}
}

@ARTICLE{1967ApJ...147..624I,
       author = {{Iben}, Jr., Icko},
        title = "{Stellar Evolution.VI. Evolution from the Main Sequence to the Red-Giant Branch for Stars of Mass 1 M\_\{sun\}, 1.25 M\_\{sun\}, and 1.5 M\_\{sun\}}",
      journal = {\apj},
         year = 1967,
        month = feb,
       volume = {147},
        pages = {624},
          doi = {10.1086/149040},
       adsurl = {https://ui.adsabs.harvard.edu/abs/1967ApJ...147..624I}
}

@ARTICLE{1971ApJ...164..111C,
       author = {{Cameron}, A.~G.~W. and {Fowler}, W.~A.},
        title = "{Lithium and the s-PROCESS in Red-Giant Stars}",
      journal = {\apj},
         year = 1971,
        month = feb,
       volume = {164},
        pages = {111},
          doi = {10.1086/150821},
       adsurl = {https://ui.adsabs.harvard.edu/abs/1971ApJ...164..111C}
}

@ARTICLE{2009ApJ...696.1823D,
       author = {{Denissenkov}, Pavel A. and {Pinsonneault}, Marc and {MacGregor}, Keith B.},
        title = "{Magneto-Thermohaline Mixing in Red Giants}",
      journal = {\apj},
         year = 2009,
        month = may,
       volume = {696},
       number = {2},
        pages = {1823-1833},
          doi = {10.1088/0004-637X/696/2/1823},
archivePrefix = {arXiv},
       eprint = {0806.4346},
 primaryClass = {astro-ph},
       adsurl = {https://ui.adsabs.harvard.edu/abs/2009ApJ...696.1823D}
}

@ARTICLE{1999MNRAS.308.1133S,
       author = {{Siess}, Lionel and {Livio}, Mario},
        title = "{The accretion of brown dwarfs and planets by giant stars - II. Solar-mass stars on the red giant branch}",
      journal = {\mnras},
         year = 1999,
        month = oct,
       volume = {308},
       number = {4},
        pages = {1133-1149},
          doi = {10.1046/j.1365-8711.1999.02784.x},
archivePrefix = {arXiv},
       eprint = {astro-ph/9905235},
 primaryClass = {astro-ph},
       adsurl = {https://ui.adsabs.harvard.edu/abs/1999MNRAS.308.1133S}
}

@ARTICLE{2019ApJ...880..125C,
       author = {{Casey}, Andrew R. and {Ho}, Anna Y.~Q. and {Ness}, Melissa and {Hogg}, David W. and {Rix}, Hans-Walter and {Angelou}, George C. and {Hekker}, Saskia and {Tout}, Christopher A. and {Lattanzio}, John C. and {Karakas}, Amanda I. and {Woods}, Tyrone E. and {Price-Whelan}, Adrian M. and {Schlaufman}, Kevin C.},
        title = "{Tidal Interactions between Binary Stars Can Drive Lithium Production in Low-mass Red Giants}",
      journal = {\apj},
         year = 2019,
        month = aug,
       volume = {880},
       number = {2},
          eid = {125},
        pages = {125},
          doi = {10.3847/1538-4357/ab27bf},
archivePrefix = {arXiv},
       eprint = {1902.04102},
 primaryClass = {astro-ph.SR},
       adsurl = {https://ui.adsabs.harvard.edu/abs/2019ApJ...880..125C}
}

@ARTICLE{2016A&A...587A..66D,
       author = {{Delgado Mena}, E. and {Tsantaki}, M. and {Sousa}, S.~G. and {Kunitomo}, M. and {Adibekyan}, V. and {Zaworska}, P. and {Santos}, N.~C. and {Israelian}, G. and {Lovis}, C.},
        title = "{Searching for Li-rich giants in a sample of 12 open clusters. Li enhancement in two stars with substellar companions}",
      journal = {\aap},
         year = 2016,
        month = mar,
       volume = {587},
          eid = {A66},
        pages = {A66},
          doi = {10.1051/0004-6361/201527196},
archivePrefix = {arXiv},
       eprint = {1512.05296},
 primaryClass = {astro-ph.SR},
       adsurl = {https://ui.adsabs.harvard.edu/abs/2016A&A...587A..66D}
}

@ARTICLE{1995ApJS..101..181W,
       author = {{Woosley}, S.~E. and {Weaver}, Thomas A.},
        title = "{The Evolution and Explosion of Massive Stars. II. Explosive Hydrodynamics and Nucleosynthesis}",
      journal = {\apjs},
         year = 1995,
        month = nov,
       volume = {101},
        pages = {181},
          doi = {10.1086/192237},
       adsurl = {https://ui.adsabs.harvard.edu/abs/1995ApJS..101..181W}
}

@ARTICLE{1999ApJ...510..217S,
       author = {{Sackmann}, I. -Juliana and {Boothroyd}, Arnold I.},
        title = "{Creation of $^{7}$Li and Destruction of $^{3}$He, $^{9}$Be, $^{10}$B, and $^{11}$B in Low-Mass Red Giants, Due to Deep Circulation}",
      journal = {\apj},
         year = 1999,
        month = jan,
       volume = {510},
       number = {1},
        pages = {217-231},
          doi = {10.1086/306545},
       adsurl = {https://ui.adsabs.harvard.edu/abs/1999ApJ...510..217S}
}

@ARTICLE{2020NatAs...4.1059K,
       author = {{Kumar}, Yerra Bharat and {Reddy}, Bacham E. and {Campbell}, Simon W. and {Maben}, Sunayana and {Zhao}, Gang and {Ting}, Yuan-Sen},
        title = "{Discovery of ubiquitous lithium production in low-mass stars}",
      journal = {Nature Astronomy},
         year = 2020,
        month = jul,
       volume = {4},
        pages = {1059-1063},
          doi = {10.1038/s41550-020-1139-7},
archivePrefix = {arXiv},
       eprint = {2007.07045},
 primaryClass = {astro-ph.SR},
       adsurl = {https://ui.adsabs.harvard.edu/abs/2020NatAs...4.1059K}
}

@ARTICLE{2023ApJ...944L...5M,
       author = {{Mallick}, Anohita and {Singh}, Raghubar and {Reddy}, Bacham E.},
        title = "{Lithium Abundances in Giants as a Function of Stellar Mass: Evidence for He Flash as the Source of Li Enhancement in Low-mass Giants}",
      journal = {\apjl},
         year = 2023,
        month = feb,
       volume = {944},
       number = {1},
          eid = {L5},
        pages = {L5},
          doi = {10.3847/2041-8213/acb5f6},
archivePrefix = {arXiv},
       eprint = {2301.10436},
 primaryClass = {astro-ph.SR},
       adsurl = {https://ui.adsabs.harvard.edu/abs/2023ApJ...944L...5M}
}

@ARTICLE{1994A&A...288..829Z,
       author = {{Zahn}, J. -P.},
        title = "{Rotation and lithium depletion in late-type binaries}",
      journal = {\aap},
         year = 1994,
        month = aug,
       volume = {288},
        pages = {829-841},
       adsurl = {https://ui.adsabs.harvard.edu/abs/1994A&A...288..829Z}
}


\appendix


\section{Linelist}
\label{app:append}

\begin{table}[!htbp]
  	\begin{center}
	\small
	\caption{Transitions used for abundance analysis of WOCS\,6008 star.}
	\label{tab6}
\begin{tabular}{ccccc} 
		\hline
		\hline
		 Wavelength & Ion & log\,gf & $\chi$\\
         (\AA) & & (dex) & (eV) \\
		\hline
		6707.761 & Li\,I & +0.002 & 1.848 \\ 
        5052.144 & C\,I & -1.303 & 7.685 \\  
        5380.325 & C\,I & -1.616 & 7.685 \\  
        6587.620 & C\,I & -1.003 & 8.537 \\  
        4982.814 & Na\,I & -0.962 & 2.104 \\
        5682.634 & Na\,I & -0.706 & 2.102 \\
        5688.205 & Na\,I & -0.452 & 2.104 \\
        6154.223 & Na\,I & -1.547 & 2.102 \\
        6160.747 & Na\,I & -1.246 & 2.104 \\
        5528.405 & Mg\,I & -0.498 & 4.346 \\
        5711.088 & Mg\,I & -1.724 & 4.346 \\
        6318.717 & Mg\,I & -2.103 & 5.108 \\
        6319.237 & Mg\,I & -2.324 & 5.108 \\
        6319.495 & Mg\,I & -2.803 & 5.108 \\
        6696.030 & Al\,I & -1.569 & 3.143 \\
        6698.662 & Al\,I & -1.870 & 3.143 \\
        6795.341 & Y\,II & -1.441 & 1.721 \\
        6795.416 & Y\,II & -1.061 & 1.738 \\
        4883.682 & Y\,II & 0.070  & 1.084 \\
        4900.119 & Y\,II & -0.090 & 1.033 \\
        4982.130 & Y\,II & -1.290 & 1.033 \\
        5087.420 & Y\,II & -0.170 & 1.084 \\
        5200.410 & Y\,II & -0.570 & 0.992 \\
        5402.775 & Y\,II & -0.360 & 1.839 \\
        5544.612 & Y\,II & -1.090 & 1.738 \\
        5546.010 & Y\,II & -1.100 & 1.748 \\
        4815.621 & Zr\,I & -0.030 & 0.604 \\
        5112.270 & Zr\,II & -0.850 & 1.665 \\
        6127.475 & Zr\,I & -1.060 & 0.154 \\
        5853.675 & Ba\,II & -0.908 & 0.604 \\
        6141.713 & Ba\,II & -0.032 & 0.704 \\
        5290.832 & La\,II & -1.750 & 0.000 \\
        5303.514 & La\,II & -1.430 & 0.321 \\
        6262.422 & La\,II & -1.240 & 0.403 \\
        6320.376 & La\,II & -1.610 & 0.173 \\
        6390.468 & La\,II & -1.450 & 0.321 \\
        6043.373 & Ce\,II & -0.480 & 1.206 \\
		\hline 
	\end{tabular}
	\end{center}
\end{table} 

\section{Composite SED fit of WOCS\,6008 with an active M-dwarf companion}
\label{app:appB}
\begin{figure}
    \centering \includegraphics[width=\columnwidth]{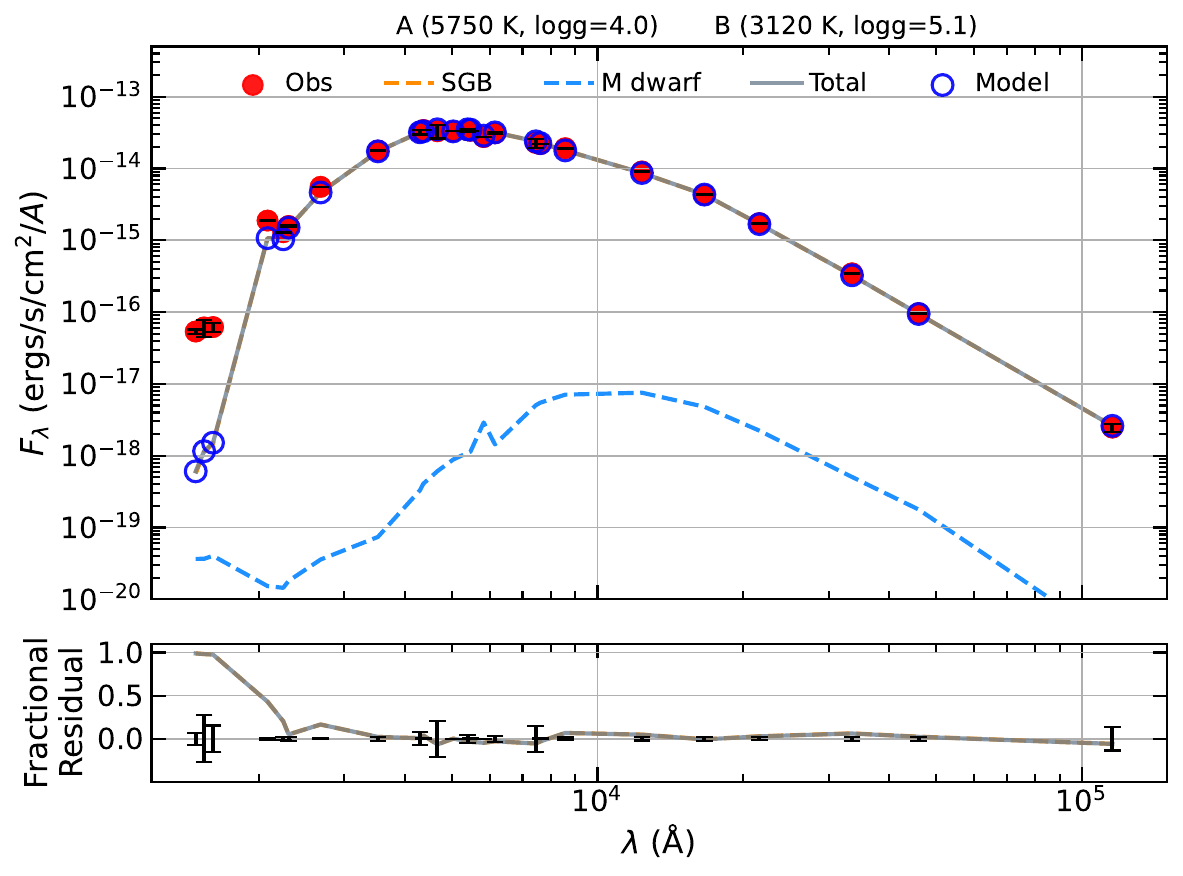}

    \caption{SED of WOCS\,6008 fitted with a combined Kurucz (orange dash-dotted) and an active M dwarf (light blue dash-dotted) observed spectrum taken from the MUSCLES survey \citep{2025ApJ...978...85W}; composite fit shown with a gray solid line. All other details are the same as in Figure~\ref{sed_fit}.}
    \label{fig:sed_mdwarf}
\end{figure}

\cite{2025ApJ...978...85W} presented observed SEDs of 12 M dwarfs with spectral types M0–M8, spanning X-ray to IR wavelengths. They observed each target long enough to capture at least one flare and did not remove the effects of stellar activity from the SEDs. To test whether an active M dwarf could account for the UV excess observed in WOCS\,6008, we selected several M dwarf spectra corresponding to the highest activity phase available from the MUSCLES survey on MAST\footnote{\url{https://archive.stsci.edu/prepds/muscles/}}, and fitted the observed SED of WOCS\,6008 with a combination of an observed active M dwarf spectrum and a Kurucz model spectrum, as an example shown in Figure~\ref{fig:sed_mdwarf}. Further details can be found in Sect.~\ref{sec:activity}.

\end{document}